\documentclass[twocolumn]{aastex631mod}

\usepackage{comment}
\usepackage{multirow}
\usepackage{booktabs}
\usepackage{amsmath,amssymb}
\usepackage{savesym}
\savesymbol{tablenum}
\usepackage{siunitx}
\restoresymbol{SIX}{tablenum}
\usepackage{hyperref}
\usepackage[nameinlink,noabbrev]{cleveref}
\usepackage[normalem]{ulem}

\makeatletter
\usepackage{etoolbox}
\patchcmd\H@refstepcounter{\protected@edef}{\protected@xdef}{}{}
\makeatother

\definecolor{Green}{HTML}{00A64F}

\newcommand*{\diff}{\ensuremath{{\rm d}}}

\shorttitle{\textit{Cosmology with BOSS void counts}}
\shortauthors{Contarini et al.}
\begin{document}

\title{Cosmological constraints from the BOSS DR12 void size function}

\correspondingauthor{Sofia Contarini}
\email{sofia.contarini3@unibo.it}

\author[0000-0002-9843-723X]{Sofia Contarini}
\affil{Dipartimento di Fisica e Astronomia ``Augusto Righi" - Alma Mater Studiorum Universit\`{a} di Bologna, via Piero Gobetti 93/2, I-40129 Bologna, Italy}
\affil{INFN-Sezione di Bologna, Viale Berti Pichat 6/2, I-40127 Bologna, Italy}
\affil{INAF-Osservatorio di Astrofisica e Scienza dello Spazio di Bologna, Via Piero Gobetti 93/3, I-40129 Bologna, Italy}
\author[0000-0002-6146-4437]{Alice Pisani}
\affil{Department of Astrophysical Sciences, Peyton Hall, Princeton University, Princeton, NJ 08544, USA}
\affil{Center for Computational Astrophysics, Flatiron Institute, 162 5th Avenue, 10010, New York, NY, USA}
\affil{The Cooper Union for the Advancement of Science and Art, 41 Cooper Square, New York, NY 10003, USA}
\author[0000-0002-0876-2101]{Nico Hamaus}
\affil{Universit\"ats-Sternwarte M\"unchen, Fakult\"at f\"ur Physik, Ludwig-Maximilians-Universit\"at, Scheinerstrasse 1, 81679 M\"unchen, Germany}
\author[0000-0002-8850-0303]{Federico Marulli$^{1,2,3}$}
\author[0000-0002-3473-6716]{Lauro Moscardini$^{1,2,3}$}
\author[0000-0003-4145-1943]{Marco Baldi$^{1,2,3}$}

%% Note that the \and command from previous versions of AASTeX is now
%% depreciated in this version as it is no longer necessary. AASTeX 
%% automatically takes care of all commas and "and"s between authors names.

%% AASTeX 6.2 has the new \collaboration and \nocollaboration commands to
%% provide the collaboration status of a group of authors. These commands 
%% can be used either before or after the list of corresponding authors. The
%% argument for \collaboration is the collaboration identifier. Authors are
%% encouraged to surround collaboration identifiers with ()s. The 
%% \nocollaboration command takes no argument and exists to indicate that
%% the nearby authors are not part of surrounding collaborations.

%% Mark off the abstract in the ``abstract'' environment. 
\begin{abstract}
We present the first cosmological constraints derived from the analysis of the void size function. This work relies on the final BOSS DR12 data set, a large spectroscopic galaxy catalog, ideal for the identification of cosmic voids. We extract a sample of voids from the distribution of galaxies and we apply a cleaning procedure aimed at reaching high levels of purity and completeness. We model the void size function by means of an extension of the popular volume-conserving model, based on two additional nuisance parameters. Relying on mock catalogs specifically designed to reproduce the BOSS DR12 galaxy sample, we calibrate the extended size function model parameters and validate the methodology. We then apply a Bayesian analysis to constrain the $\Lambda$CDM model and one of its simplest extensions, featuring a constant dark energy equation of state parameter, $w$. Following a conservative approach, we put constraints on the total matter density parameter and the amplitude of density fluctuations, finding $\Omega_{\rm m}=0.29 \pm 0.06$ and $\sigma_8=0.79^{+0.09}_{-0.08}$. Testing the alternative scenario, we derive $w=-1.1\pm 0.2$, in agreement with the $\Lambda$CDM model. These results are independent and complementary to those derived from standard cosmological probes, opening up new ways to identify the origin of potential tensions in the current cosmological paradigm.

\end{abstract}

%% Keywords should appear after the \end{abstract} command. 
%% See the online documentation for the full list of available subject
%% keywords and the rules for their use.

%% From the front matter, we move on to the body of the paper.
%% Sections are demarcated by \section and \subsection, respectively.
%% Observe the use of the LaTeX \label
%% command after the \subsection to give a symbolic KEY to the
%% subsection for cross-referencing in a \ref command.
%% You can use LaTeX's \ref and \label commands to keep track of
%% cross-references to sections, equations, tables, and figures.
%% That way, if you change the order of any elements, LaTeX will
%% automatically renumber them.
%%
%% We recommend that authors also use the natbib \citep
%% and \citet commands to identify citations.  The citations are
%% tied to the reference list via symbolic KEYs. The KEY corresponds
%% to the KEY in the \bibitem in the reference list below. 

\section{Introduction} \label{sec:intro}

Over the past decade, cosmology has witnessed an unprecedented technological advancement prompting the release of a huge amount of data about our Universe. This has been made possible thanks to a massive experimental effort conducted by means of wide and deep-field surveys, complemented with the development of large and complex cosmological simulations. In this context, the study of cosmic voids, vast underdense regions of our Universe, began to provide its fundamental contribution in addressing fundamental questions in cosmology \citep[see][for a review]{pisani2019,Moresco2022}. 

With their large extent and unique low-density interior, voids indeed represent the ideal environment to study dark energy and modified gravity theories \citep{clampittMG2013,spolyarMG2013,cai2015,pisani2015,hamaus2015,zivickMG2015,pollina2015,achitouv2016,sahlen2016,falck2018,sahlen2018,paillas2019,perico2019,verza2019,Contarini2020,Contarini2022}, as well as models including massive neutrinos \citep{massara2015,banerjee2016,kreisch2018,sahlen2019,schuster2019,Contarini2020,Kreisch2021}, primordial non-Gaussianities \citep{chan2019} and phenomena beyond the standard model of particle physics \citep[e.g.][]{reed2015,yang2015,baldi2016,Lester2021,Acari2022}.

The most straightforward analysis in this context is to count cosmic voids as a function of their size. This statistic is called the \textit{void size function}. In recent years it has been explored in details through cosmological simulations \citep[e.g.][]{pisani2015,pollina2017,ronconi2019,contarini2019,verza2019,Contarini2020,Contarini2022,Pelliciari} 
and more rarely with real survey catalogs \citep{sahlen2016,nadathur2016b,pollina2019}. Another fundamental and extensively studied void statistic is the void-galaxy cross-correlation function, which represents an equivalent of the void density profile \citep{hamaus2014,Schuster2022}. It has been exploited for cosmological purposes by means of modeling the geometric and dynamic effects on the average void shape \citep{lavaux2012,pisani2014,hamaus2016,cai2016,hawken2017,hamaus2017,achitouv2019,correa2019,hamaus2020,hawken2020,aubert2020,nadathur2020,Correa2021,EuclidHamaus2021,nadathur2022,woodfinden2022}. Other notable and modern void analyses concern the weak lensing phenomenon \citep{clampitt2015,chantavat2016,gruen2016,chantavat2017,cai2017,sanchez2017,baker2018,brouwer2018,fang_DES2019,vielzeuf2019,Davis2021,Bonici2022}, as well as the void auto-correlation function \citep{hamaus2014b,Hamaus2014c,chan2014,clampitt2016,chuang2017,kreisch2018,voivodic2020}, and the impact from voids on the integrated Sachs–Wolfe effect in the cosmic microwave background \citep{granett2008,cai2014,cai2017,kovacs2018,kovacs2019,vielzeuf2019,dong2020,hang_2021,kovacs2022}.

The various void statistics allow us to investigate the theoretical foundations of our Universe and test the current standard cosmological scenario, i.e. the $\Lambda$-cold dark matter ($\Lambda$CDM) model. Using real data catalogs to obtain effective cosmological constraints on the $\Lambda$CDM parameters, or to examine its main alternatives \citep[see e.g.][for a review]{Yoo2012,Joyce2016,Amendola2018}, requires complete and statistically relevant void samples.  
For this reason, cosmic voids are usually analyzed in spectroscopic wide-field surveys \citep[e.g.,][]{sutter2012a,nadathur2016b,hamaus2016,mao2017,hawken2020,kovacs2021}, covering large and possibly contiguous regions of the sky \citep[see][for tests on the impact of the survey mask on void statistics]{sutter2014b}.

On the basis of this requirement, we here employed the final data release (DR12) of the Baryon Oscillation Spectroscopic Survey (BOSS, \citealp{BOSSDR12}), provided by the Sloan Digital Sky Survey (SDSS-III, \citealp{SDSS2011}). Thanks to its wide and contiguous sky area coverage, this galaxy catalog is indeed particularly suited to study cosmic voids. \cite{mao2016} explored the void sample extracted from the distribution of the BOSS DR12 galaxies and assessed its purity taking into account survey boundaries, masks, and radial selections. Then, assuming stacked voids to be spherically symmetric, \cite{mao2017} performed the Alcock-Paczyński test \citep{AP1979} to the same sample of voids in order to evaluate the matter content of the Universe, $\Omega_{\rm m}$.
Moreover, \cite{hamaus2016} made use of BOSS data to study the joint effects of geometric and dynamic distortions on the stacked void-galaxy cross-correlation function, deriving precision-level constraints on $\Omega_{\rm m}$ and on the growth rate of structures. Other important works focusing on the study of voids from the SDSS have been conducted during recent years \citep[e.g.,][]{hoyle2004,pan2012,sutter2012b,sutter2014b,mao2016,hamaus2017,hamaus2020,nadathur2020,aubert2020}, but in none of these the void size function has been exploited as a cosmological probe.

In this paper, our goal is to measure and model the size function of BOSS DR12 voids to test the standard cosmological scenario. Since it offers cosmological constraints that are complementary to those of standard cosmological probes, this kind of analysis is extremely promising \citep{Bayer2021,Kreisch2021,Contarini2022,Pelliciari}. In particular, the negligible correlation between void counts and the main overdensity-based statistics (e.g., galaxy clustering and the halo mass function), together with the striking orthogonality of the corresponding growth of structure constraints (i.e., the $\Omega_{\rm m}$--$\sigma_8$ confidence contours), has revealed the utmost importance of modeling the void size function, in the perspective of combining its achievements with those of other cosmological probes.

The analysis that we present in this paper represents the first cosmological exploitation of the void size function as stand-alone probe \citep[see however][for the modeling of extreme-value void count statistics]{sahlen2016}.
With this pioneering work, we provide valuable cosmological constraints on the $\Omega_{\rm m}$--$\sigma_8$ parameter plane and we test the standard $\Lambda$CDM scenario by analysing its simplest alternative based on a constant dark energy equation of state.

The paper is organised as follows. In \Cref{sec:theory} we provide the theoretical foundations of the void size function model, and extend its standard theory to include observational effects and parametrise the galaxy bias contribution on the void radius. In \Cref{sec:analysis} we describe the galaxy catalogs employed for the analysis, together with the procedure of void identification and cleaning. Moreover, we outline the details of the Bayesian analysis performed to test different cosmological scenarios and we present the results of the void size function model calibration, attained through void mock catalogs. In \Cref{sec:results} we apply the calibrated void size function model to the BOSS DR12 data and present the main outcome of the work. Finally, in \Cref{sec:conclusions} we draw our conclusions and discuss future applications of the analysis.

\section{Theory} \label{sec:theory}

\subsection{Void size function}\label{sec:VSF_theory}
The comoving number counts of voids as a function of their radius, i.e. the void size function, has been modeled from first principles by \cite{SVdW2004}. This model is developed within the framework of the excursion-set theory, with the same formalism adopted for the halo mass function \citep{PressSchechter,peacock_heavens1990,cole1991,bond1991,mo_white1996}. 
The multiplicity function, expressing in this case the volume fraction of the Universe occupied by cosmic voids in linear theory\footnote{All the quantities computed in linear theory are indicated in this work by the superscript ``L''. We will also mark the nonlinear ones with the superscript ``NL'' when it is necessary to distinguish.}, yields:
\begin{equation}\label{eq:multiplicity}
f_{\ln \sigma_{\rm m}}(\sigma_{\rm m}) = 2 \sum_{j=1}^{\infty} \, \exp{\bigg(-\frac{(j \pi x)^2}{2}\bigg)} \, j \pi x^2 \, \sin{\left( j \pi \mathcal{D} \right)}\, ,
\end{equation}
with
\begin{equation}\label{eq:deltas}
\mathcal{D} = \frac{|\delta_\mathrm{v}^\mathrm{L}|}{\delta_\mathrm{c}^\mathrm{L} + |\delta_\mathrm{v}^\mathrm{L}|}\, , \qquad x = \frac{\mathcal{D}}{|\delta_\mathrm{v}^\mathrm{L}|} \sigma_{\rm m} \, ,
\end{equation}
where $\sigma_{\rm m}$ is the root mean square variance of the linear matter perturbations on a scale $R_\mathrm{L}$, while $\delta_\mathrm{v}^\mathrm{L}$ and $\delta_\mathrm{c}^\mathrm{L}$ are the two fundamental barriers of the model\footnote{$\delta_\mathrm{v}^\mathrm{L}$ is the linear density contrast associated with the formation of voids, while $\delta_\mathrm{c}^\mathrm{L}$ with the collapse of dark matter halos. The value of the former is $\delta_\mathrm{v}^\mathrm{L}=-2.71$ in an Einstein--de Sitter scenario; this value, however, cannot be directly used to model voids identified in a realistic tracer field \citep{contarini2019} and should be properly corrected to match the properties of the selected void sample (see details in \Cref{sec:cleaning}). The latter can generally vary in the range $ 1.06 \le \delta_\mathrm{c}^\mathrm{L} \le 1.686$, given by the turn-around and the collapse linear thresholds in an Einstein--de Sitter model. Nevertheless, we note that the variation of the quantity $\delta_\mathrm{c}^\mathrm{L}$ affects only the number of very small voids, which are not included in our analysis.}.

The multiplicity function of \Cref{eq:multiplicity} assumes initial spherical density fluctuations to evolve according to linear theory over cosmic time. Therefore, to predict the number counts of observed voids of size $R$, we need to consider the nonlinear evolved density field.
For this purpose, \cite{jennings2013} modified the void size function model imposing the conservation of the volume occupied by voids during the transition from linearity to nonlinearity, which for this reason is called \textit{volume-conserving} (hereafter, Vdn) model and is expressed as:
\begin{equation}
\frac{\diff n}{\diff \ln R} = \frac{f_{\ln \sigma_{\rm m}}(\sigma_{\rm m})}{V(R)} \, \frac{\diff \ln \sigma_{\rm m}^{-1}}{\diff \ln R_\mathrm{L}} \biggr \rvert_{R_\mathrm{L} = R_\mathrm{L}(R)}\, .
\end{equation}
The Vdn model successfully predicts the size function of cosmic voids identified in the total matter field (i.e. in the distribution of dark matter particles), provided that these voids are defined as spherical non-overlapping spheres embedding a specific density contrast $\delta_\mathrm{v, DM}^\mathrm{NL}$ \citep{jennings2013, ronconi2017, ronconi2019, contarini2019}. The threshold $\delta_\mathrm{v, DM}^\mathrm{NL}$ is the nonlinear counterpart of the quantity $\delta_\mathrm{v}^\mathrm{L}$ introduced in \Cref{eq:deltas}, where the subscript ``DM'' specifies the requirement to be computed in the dark matter density field. To convert the nonlinear threshold into its linear counterpart and vice-versa, we can make use of the fitting formula provided by \cite{bernardeau1994}:
\begin{equation} \label{eq:bernardeau}
  \delta_\mathrm{v}^\mathrm{L} = \mathcal{C} \, \bigl[1 - (1 + \delta_\mathrm{v}^\mathrm{NL})^{-1/\mathcal{C}}\bigr] \ , \ \text{with } \mathcal{C}=1.594 \, .
\end{equation}
which features enhanced precision when applied to underdensities.

One of the observational effects to consider when computing the theoretical void size function are the so-called geometric distortions, which are caused by the assumption of a fiducial cosmology that may not coincide with the true one. Geometric distortions affect the observed void shape by introducing an anisotropy between the directions parallel and perpendicular to the line of sight. This distortion is commonly denoted as the \cite{AP1979} (AP) effect and can be modeled following the prescriptions by, e.g., \cite{Sanchez2017b}:
\begin{equation}
\begin{gathered}
r_\parallel = \frac{H^*(z)}{H(z)} \, r^*_\parallel \equiv q_\parallel \, r^*_\parallel \,, \\
r_\perp = \frac{D_{\rm A}(z)}{D^*_{\rm A}(z)} \, r^*_\perp \equiv q_\perp \, r^*_\perp \, .
\end{gathered}
\end{equation}
where $r^*_\parallel$ and $r^*_\perp$ are the observed comoving distances between two objects at redshift $z$, projected along the parallel and perpendicular directions, respectively, $H(z)$ is the Hubble parameter and $D_{\rm A}(z)$ is the comoving angular-diameter distance. The quantities marked with asterisks are those computed by assuming the fiducial cosmology, while the remaining ones are calculated in the true cosmology. In the case of cosmic voids we can derive the radius $R$ of a sphere having the same volume of the observed deformed void as $R=q_\parallel^{1/3}q_\perp^{2/3} R^*$ \citep{Ballinger1996, Eisenstein2005_BAO, xu2013, Sanchez2017b, hamaus2020, Correa2020}. The application of this correction to the void size function model translates into a shift of the predicted void counts towards smaller or larger radii, depending on the parameter values assumed for the fiducial cosmology.

\subsection{Vdn model parametrization}\label{sec:Vdn_extension}

When dealing with real survey data, cosmic voids cannot be defined in the dark matter particle field, so an extension of this modeling to biased tracers, such as galaxies and galaxy clusters, is required. In particular, it is fundamental to consider the effect of tracer bias on the void density profiles and consequently on the associated void radius.

The density profiles of voids traced by biased objects are deeper than those traced by dark matter particles. The radius corresponding to a fixed internal spherical density contrast is consequently larger for the former \citep[e.g.,][]{Sutter2013,pollina2017,Schuster2022}. To include this effect in the void size function theory, \cite{contarini2019} proposed a parametrization of the Vdn model nonlinear underdensity threshold as a function of the tracer effective bias, $b_\mathrm{eff}$:
\begin{equation}\label{eq:thr_conversion}
\begin{aligned}
 \delta_\mathrm{v,DM}^\mathrm{NL} &= \frac{\delta_\mathrm{v,tr}^\mathrm{NL}}{\mathcal{F}(b_\mathrm{eff},z)}\, , \ \text{with} \\
 \mathcal{F}(b_\mathrm{eff},z) &= B_\mathrm{slope} \, b_\mathrm{eff}(z) + B_\mathrm{offset}\, ,
\end{aligned}
\end{equation}
where $\delta_\mathrm{v,tr}^\mathrm{NL}$ is the tracer density contrast embedded by voids identified in the tracer field and, as already mentioned, $\delta_\mathrm{v,DM}^\mathrm{NL}$ is the matter underdensity threshold to be used in the void size function model, after its conversion from linear theory. The two coefficients regulating the slope and the offset of the linear function $\mathcal{F}(b_\mathrm{eff})$, i.e. $B_{\rm slope}$ and $B_{\rm offset}$, are redshift independent and can be derived by calibrating the void size function model with simulations \citep{contarini2019,Contarini2020,Contarini2022}. They have been shown to depend on the kind of tracers used to identify the voids, but also to have a negligible dependence on the cosmological model \citep{Contarini2020}.

Despite its simplicity and demonstrated effectiveness, this model parametrization requires an accurate measurement of the tracer effective bias, $b_\mathrm{eff}$. This quantity is not trivial to estimate in real data catalogs, since it exhibits a degeneracy with the matter power spectrum normalization, $\sigma_8$. For example, by modeling the two-point correlation function (2PCF), only the product of these parameters, $b_\mathrm{eff} \, \sigma_8$, can be derived. Therefore, the assumption of a fiducial value of $\sigma_8$ is necessary to recover separately the galaxy effective bias. To avoid this problem, in this work we introduce a different parametrization of the linear function $\mathcal{F}$, which also depends on $\sigma_8$:
\begin{equation}\label{eq:new_F}
\mathcal{F}(b_\mathrm{eff} \, \sigma_8, z) = C_\mathrm{slope} \, b_\mathrm{eff}(z) \, \sigma_8(z)+ C_\mathrm{offset}\, .
\end{equation}
This expression has the advantage to be based on a $\sigma_8$-dependent bias, i.e. the product $b_\mathrm{eff} \, \sigma_8$, which can be estimated directly via the modeling of the 2PCF.

The linear parametrization reported in \Cref{eq:thr_conversion} and equivalently the form of \Cref{eq:new_F} is effective also to encapsulate redshift-space distortions \citep{Contarini2022}, i.e. the deformation of void shapes due to the peculiar velocities of tracers. Indeed, when estimating comoving distances from redshifts, the coherent outflow of tracers around the void center leads to larger observed void radii on average \citep{pisani2015b}.

Therefore, in this work we model the void size function by means of the Vdn theory modified to correct for both geometric and dynamic effects, relying on an underdensity threshold parametrization dependent on the quantity $b_\mathrm{eff} \, \sigma_8$. Hereafter, we will refer to this model as \textit{extended} Vdn.

\section{Analysis} \label{sec:analysis}

\subsection{Data preparation} \label{sec:data_prep}

In this work we analyze the BOSS DR12 galaxy sample, covering both the northern and the southern galactic hemispheres, for a total sky area of more than ten thousand square degrees. We account for the two target selections denoted LOWZ and CMASS \citep{Reid2016}, whose combination features more than one million galaxies spanning the redshift range $0.2<z<0.75$ and constitutes one of largest spectroscopic catalogs available up to date. 

We also make use of the MultiDark PATCHY mock catalogs, specifically designed to mimic relevant observational effects characterizing the BOSS DR12 galaxy sample, including selection functions and masking. In particular, we consider $100$ different realizations of these mocks to calibrate and validate our void size function model. These catalogs have been produced by populating the dark matter halos of the MultiDark cosmological simulations \citep{MultiDark2016} with the PATCHY method \citep{kitaura2014}, which accurately reproduces the clustering properties of different populations of objects \citep{BOSSDR12_clustering}. In particular, \cite{Kitaura2016} and \cite{rodriguez2016} demonstrated that the PATCHY mocks follow accurately the structure formation down to scales of a few megaparsec, as well as the evolution of galaxy bias and nonlinear redshift-space distortions. Moreover, PATCHY mocks reproduce the BOSS DR12 power spectrum, two- and three-point correlation functions up to $k \sim 0.3 \ h \ \mathrm{Mpc}^{-1}$, for arbitrary stellar mass bins.

The MultiDark simulation has been run with a flat $\Lambda$CDM cosmology characterized by Planck2013 parameters \citep{Planck2013}: ${\Omega_{\rm m}=0.307115}$, $\Omega_{\rm b}=0.048206$, $\sigma_8=0.8288$, $n_{\rm s}=0.9611$ and $h=0.6777$. Although many mock realizations are available, all the PATCHY simulations are built with this set of cosmological parameters. We do not expect this to impact significantly our results, despite our aim at analyzing cosmological models with different values of $\Omega_\mathrm{m}$, $\sigma_8$ and $w$.
Indeed, the effectiveness of the void size function model has been proved across various cosmological parameter ranges and alternative cosmological scenarios \citep[see e.g.][]{verza2019,Contarini2020}.

\begin{figure*}
\centering
\includegraphics[width=0.363\textwidth]{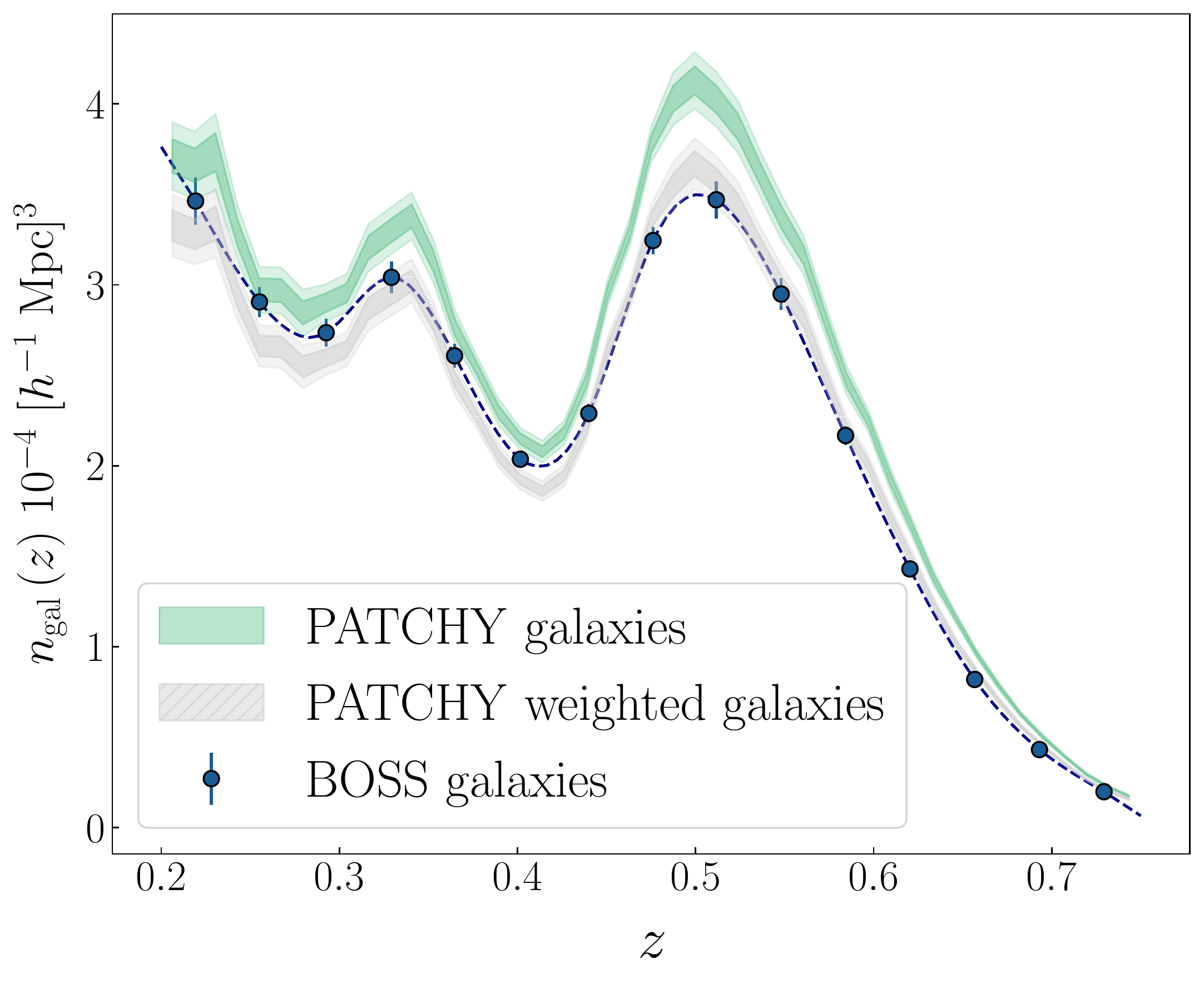}
\includegraphics[width=0.63\textwidth]{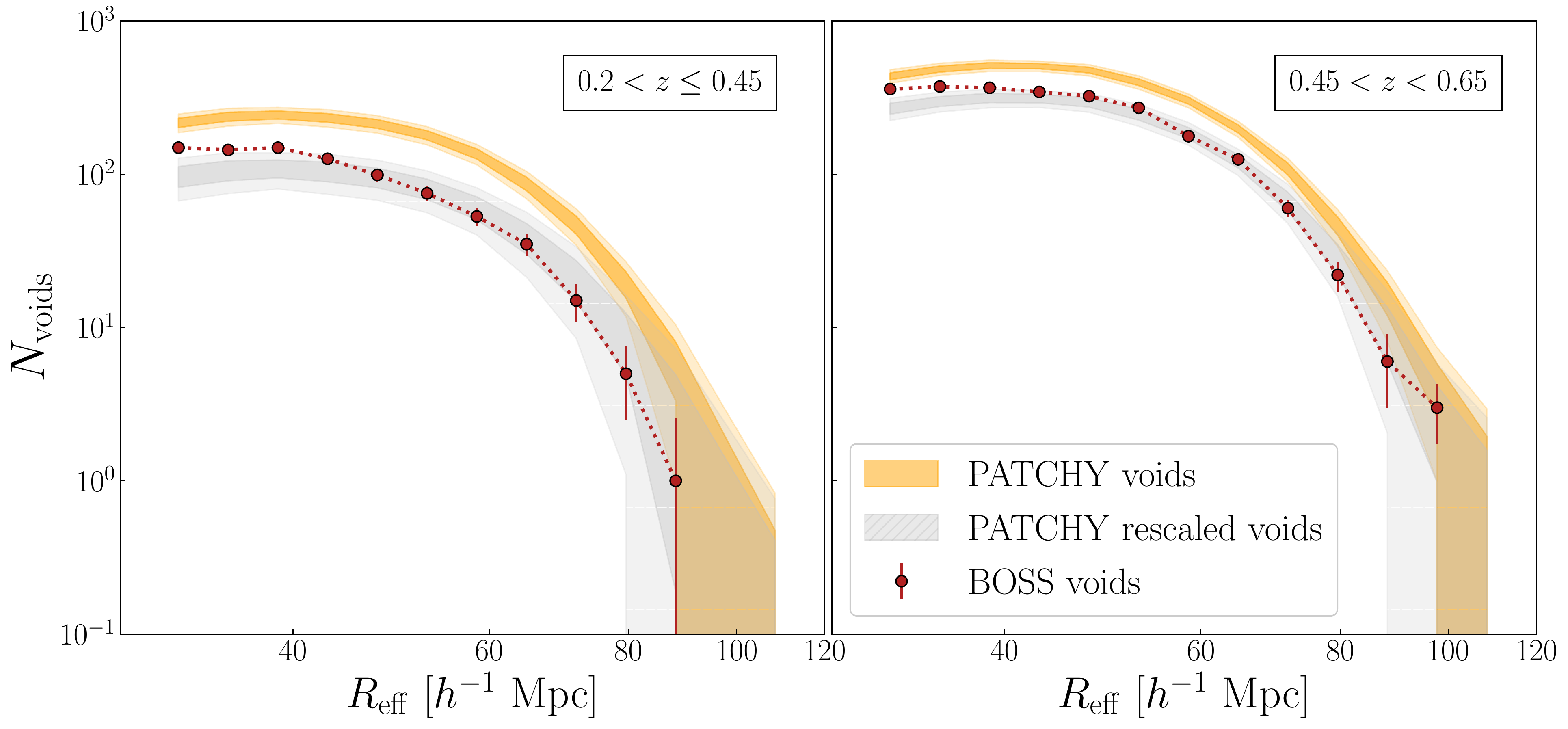}
\caption{\textit{Left} panel: number density of the BOSS DR12 and MultiDark PATCHY galaxies as a function of redshift. The BOSS data are represented with blue markers and a spline fit in dashed. We report with a green band the galaxy number density of the $100$ PATCHY mock realizations, distinguishing with different shades the $68\%$ and $95\%$ confidence levels. We represent the same distribution, but for mock galaxies weighted to mimic the BOSS survey completeness in grey. \textit{Central} and \textit{right} panels: number counts measured from the full sample of voids identified with \texttt{VIDE}, computed as a function of their effective radius and with a cut $R_\mathrm{eff} > 30 \ h^{-1} \ \mathrm{Mpc}$. The two panels represent the void size function in the redshift ranges $0.2 < z \leq 0.45$ and $0.45 < z < 0.65$., in red circles with error bars for the BOSS DR12 data and in orange bands for the MultiDark PATCHY mocks. The differently shaded bands show the $68\%$ and $95\%$ confidence levels around the mean value of the PATCHY void number counts. Grey bands represent PATCHY void counts after rescaling to the slightly smaller volume of the BOSS catalog. \label{fig:VIDE_VSF}}
\vspace{0.2cm}
\end{figure*}
We identify the cosmic underdensities in the distribution of galaxies both in real and simulated data by means of the public Void IDentification and Examination toolkit\footnote{\url{https://bitbucket.org/cosmicvoids/vide_public}} \citep[\texttt{VIDE},][]{vide}. \texttt{VIDE} is a parameter-free void finding algorithm based on the code ZOnes Bordering On Voidness \citep[\texttt{ZOBOV},][]{zobov}, which performs a Voronoi tessellation of tracer particles to reconstruct a continuous density field. Then it finds the local density minima, from which the Voronoi cells start to be merged progressively to create density basins through a watershed algorithm \citep{platen2007}. The whole 3D space is thereby divided up by thin overdense ridges into separate underdense regions, forming the final catalog of voids. For each underdensity identified, the corresponding void center position $\mathbf{X}$ is calculated as the volume-weighted barycenter of the $N$ Voronoi cells that define the void, while the void effective radius $R_{\rm eff}$ is taken as the radius of a sphere with the total volume of all those cells:
\begin{equation}
    \mathbf{X} \equiv \frac{\sum_{i=1}^{N}{\mathbf{x}_i V_i}}{\sum_{i=1}^{N}{V_i}} \quad \textrm{ and} \quad R_{\rm eff} \equiv \left(\frac{3}{4\pi} \sum_{i=1}^{N}{V_i}  \right)^{1/3} \textrm{ ,}
\end{equation} 
where $\mathbf{x}_i$ are the coordinates of the $i$-th tracer of a given void, and $V_i$ is the volume of its associated Voronoi cell.

We apply \texttt{VIDE} to the BOSS DR12 galaxy sample and to the $100$ PATCHY realizations, assuming throughout the analysis the Planck2013 cosmological parameters. This cosmology is used to convert angles and redshifts into comoving coordinates to build the galaxy map, and is assumed as fiducial to compute the void size function model in the Bayesian approach we present in \Cref{sec:cosmological_modeling}. The impact of the fiducial cosmology assumption on the results presented in this paper is analyzed in \ref{appendix:systematics}.

The number of cosmic voids identified by the algorithm is strictly related to both the survey volume and the number density of the objects used to trace voids \citep[see e.g.,][]{Schuster2022}. We report in the left panel of \Cref{fig:VIDE_VSF} the number density of galaxies computed in concentric redshift shells, $n_{\rm gal} (z)$, for both BOSS and PATCHY samples. We show the results for the PATCHY data set as a shaded band representing the $68\%$ and $95\%$ confidence levels around the mean number density value, computed from the standard deviation between the $100$ mock realizations. The PATCHY data standard deviation is used also to calculate the uncertainty associated with the number density of BOSS galaxies, reported in this plot with $2\sigma$ error bars for visibility, and interpolated with a spline function. As a comparison, we also show the number density distribution of mock galaxies processed by means of the weights specifically designed for the MultiDark PATCHY catalogs, which mimic observational systematic effects that affect the completeness of the galaxy sample. We underline, however, that the weighted PATCHY samples will not be used in the following analysis, since our aim is to exploit mock catalogs to calibrate the void size function model and validate our pipeline. Indeed, using galaxy catalogs characterized by a higher completeness allows us to obtain tighter constraints on the calibration parameters and does not bias the final results with BOSS data, since the galaxy clustering of the two samples are fully compatible on the relevant range of scales (see \ref{appendix:multipoles}).

As expected, the intrinsic higher mean density of the PATCHY galaxy catalog with respect to the BOSS one is reflected in the larger void numbers of these samples. The total number of voids identified with the void finder \texttt{VIDE} above a minimum effective radius of $30 \ h^{-1} \ \mathrm{Mpc}$ is $N_{\rm v}=3280$ for the BOSS DR12 data and $\overline{N}_{\rm v}=5111.85$ as a mean value for the $100$ PATCHY mock realizations. We note that in this estimate, as well as in the analysis we will present, we exclude galaxies with $z>0.65$, since their number density is very low in this regime and extended objects like voids are potentially affected by the survey boundary.

We report in the rightmost panels of \Cref{fig:VIDE_VSF} the size distribution of void number counts, dividing the sample in two bins of redshift, $0.2 < z \leq 0.45$ and $0.45 < z < 0.65$, approximately coinciding with the two target selections LOWZ and CMASS. We show the void count size distribution from PATCHY via two shaded bands representing confidence levels of $68\%$ and $95\%$ as obtained from the $100$ realizations. The corresponding uncertainty on the BOSS void number counts is computed by rescaling the PATCHY $68\%$ errors by the square root of the ratio between BOSS and PATCHY survey volumes. Moreover, we use this survey volume ratio between PATCHY mocks and BOSS data to rescale the void counts in PATCHY, as shown in the two rightmost panels of \Cref{fig:VIDE_VSF}.

We computed the exact volume for these galaxy catalogs by summing up all Voronoi cells created by \texttt{VIDE} to tessellate the 3D space. This corresponds approximately to the total volume occupied by all the voids identified by the algorithm. This approach also takes into account the removal of voids intersecting survey boundaries (see \citealp{vide} for further details). The effective volumes derived with this procedure will be used in the following analysis also to convert the void number density predictions provided by the extended Vdn theory into void counts, $N_{\rm voids}$, to model accordingly the measured void size function.

As evident from \Cref{fig:VIDE_VSF}, the size distribution of the rescaled PATCHY void counts matches the BOSS data closely. Only towards the small-scale cut, at ${R_{\rm eff}<35 \ h^{-1} \ \mathrm{Mpc}}$, small deviations occur, which may be caused by finite resolution effects close to the mean galaxy separation ${\rm MGS}(z)=n_{\rm gal}(z)^{-1/3}$ \citep[see, e.g.,][for further details]{jennings2013,pisani2015,ronconi2017,contarini2019,Contarini2022}. Following our cleaning procedure, as described in \Cref{sec:cleaning}, these small voids will be excluded in our analysis anyway. Finally, we note that the void counts from PATCHY can reach numbers below unity due to the averaging procedure over all mock catalogs.

\subsection{Void catalog cleaning} \label{sec:cleaning}

\begin{figure*}
\centering
\includegraphics[width=0.8\textwidth]{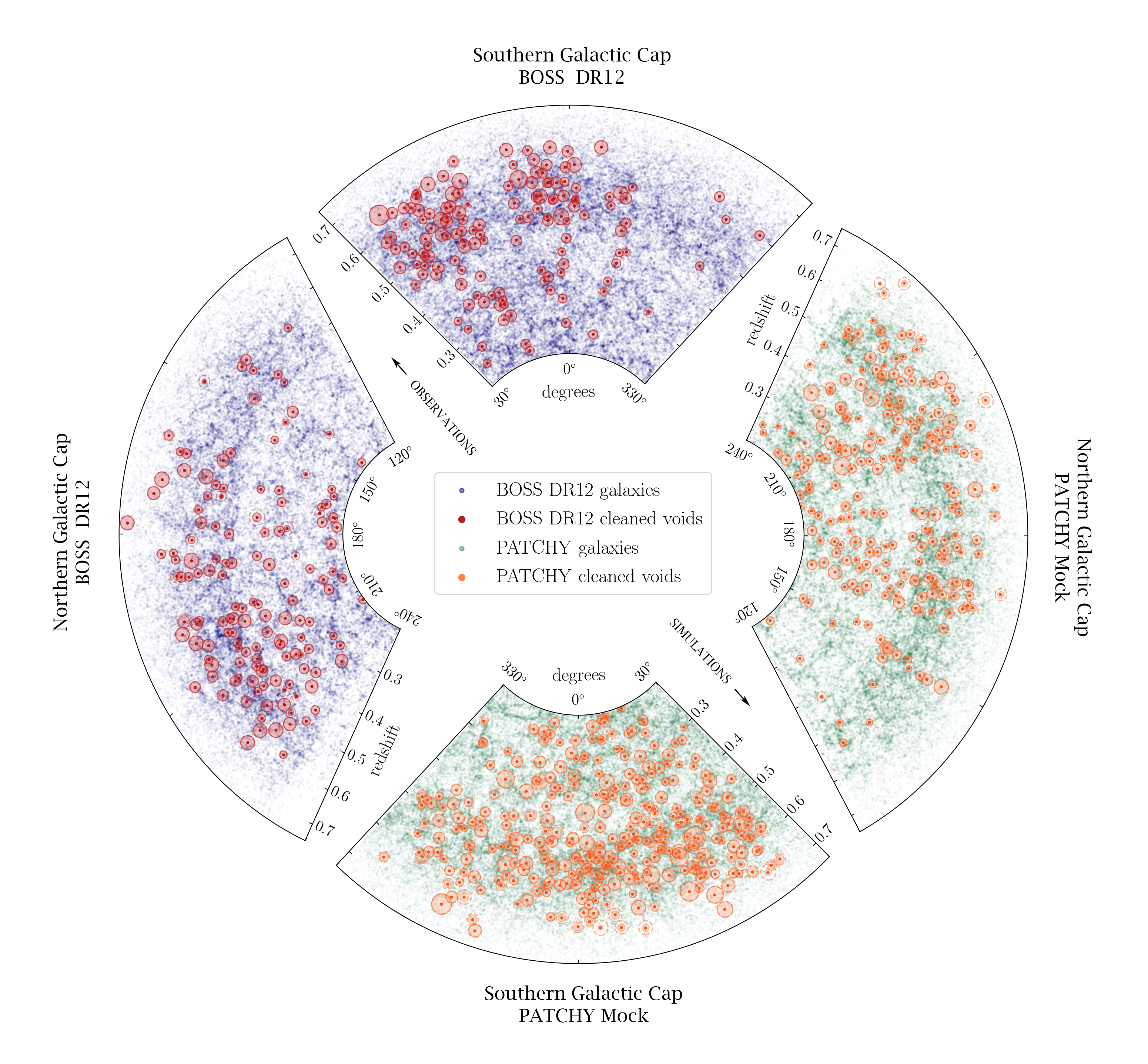}
\caption{Visual representation of the distribution of galaxies and voids in the BOSS DR12 data set and the MultiDark PATCHY simulations from a redshift range $0.2<z<0.75$ and within a slice of width $200 \ h^{-1} \ \mathrm{Mpc}$. In the \textit{upper-left} panels we show the northern and the southern galactic caps covered by BOSS, while in the \textit{lower-right} panels the corresponding regions reproduced by one of the PATCHY mocks. For each of the four represented sky areas we indicate the cosmic voids identified in the distribution of galaxies by filled circles, post-processed with the cleaning procedure. For those voids that only partially intersect with the selected slice we additionally show a dashed circle to indicate their full size that extends beyond the slice. \label{fig:visual}}
\vspace{0.2cm}
\end{figure*}

The void sample provided by \texttt{VIDE} is not yet directly exploitable for the study of the void size function \citep{jennings2013,ronconi2017}, unless a number of ad hoc correction factors are introduced in order to link the cosmic voids found by the void finder with the ideal voids considered in the void size function theory. As introduced in \Cref{sec:theory}, the Vdn model predicts the number counts of voids defined as spherical non-overlapping regions embedding a specific density contrast value in the total matter field, $\delta_\mathrm{v,DM}^\mathrm{NL}$.

In this work, we adopt the complementary approach of aligning the considered sample of voids with the prescriptions of the Vdn model \citep[as previously done by][]{ronconi2019,contarini2019,verza2019,Contarini2020,Contarini2022}, therefore without modifying the void definition used in the theoretical model. To this end, we follow the cleaning methodology proposed by \cite{jennings2013}, applying an upgraded version of the code implemented by \cite{ronconi2017}. In this algorithm, the only settable parameter is the threshold $\delta_\mathrm{v,tr}^\mathrm{NL}$, which is computed in the density field of tracers (i.e. galaxies in this case) and is used to prepare the void sample. In this upgraded version of the cleaning algorithm, the galaxy density distribution is approximated as reported in the left panel of \Cref{fig:VIDE_VSF}, i.e. by a spline fit to the number density of galaxies as a function of redshift. 

In our analysis we follow the choice of \cite{Contarini2022} to select a threshold of $\delta_\mathrm{v,tr}^\mathrm{NL}=-0.7$: this density contrast has been proven to represent a good compromise to model the innermost underdense regions of voids and simultaneously avoid poor statistics due to the sparsity of tracers. We underline that the selection of this threshold is not unique. We refer the reader to \cite{Contarini2022} for further details about the impact of this parameter choice. 

Once the threshold $\delta_\mathrm{v,tr}^\mathrm{NL}$ has been selected, the cleaning algorithm firstly discards from the original \texttt{VIDE} void catalog all those underdensities having an internal spherical density contrast excessively high: starting from the void center identified by \texttt{VIDE}\footnote{We note that relying on the \texttt{VIDE} void center includes, to some extent, knowledge about the void shape, since it is computed as barycenter of all the tracers composing the density depression (see \Cref{sec:data_prep}).}, a sphere with starting radius $R_{\rm central} = 2 \, {\rm MGS}(z)$ is grown, and if this sphere does not match the minimum requirement imposed by the chosen threshold, i.e. $\delta_\mathrm{v,tr}(R_{\rm central})<-0.7$, the corresponding void is rejected. Secondly, all the remaining underdensities are rescaled following the shape of their void density profile up to a radius $R$,\footnote{We highlight the different notation with respect to the effective radius of voids not processed by the cleaning procedure, i.e. $R_{\rm eff}$.} embedding a spherical density contrast equal to $\delta_\mathrm{v,tr}^\mathrm{NL}$. Lastly, all the mutually intersecting voids are removed by rejecting one by one the overlapping object with the higher central density.

The final outcome of this cleaning procedure is a sample of voids ideally free of spurious Poissonian fluctuations and shaped in agreement with the theoretical definition adopted in the void size function model. In particular, to match the void number counts measured from this sample and the void size function predicted by the extended Vdn model (see \Cref{sec:theory}), we must convert the underdensity threshold used to rescale the voids, $\delta_\mathrm{v,tr}^\mathrm{NL}=-0.7$ in our case, by means of \Cref{eq:new_F}; this allows us to recover the corresponding value in the total matter density field, $\delta_\mathrm{v,DM}^\mathrm{NL}$, which then is inserted in \Cref{eq:deltas} after the transformation to its linear theory counterpart via \Cref{eq:bernardeau}.

We show an example of the cleaned void samples obtained in this work in \Cref{fig:visual}, where we report $200 \ h^{-1} \ \mathrm{Mpc}$-thick slices of the BOSS DR12 and MultiDark PATCHY light-cones. In this visual representation we depict galaxies and voids in the full redshift range $0.2<z<0.75$, separating the observed BOSS data (top-left panels) from those corresponding to one of the $100$ PATCHY mocks (bottom-right panels). As expected, we see a similarity between the projected distribution of galaxies and voids in the specular sky sections of the observed and simulated data. We also point out that any apparent overlapping between voids is given by projection effects within the slice and that the regions not featuring the presence of voids are those excluded by the survey mask.

\subsection{Cosmological analysis}\label{sec:cosmological_modeling}
Now that the void size function theory and the void sample preparation have been outlined, we describe the Bayesian analysis we perform to investigate different cosmological scenarios and to calibrate the void size function model (see \Cref{sec:model_calibration}).
According to Bayesian statistics, the posterior probability of a set of model parameters, $\Theta$, given a data set, $\mathcal{D}$, which is composed of the measured void number counts in our case, can be computed as: ${\mathcal{P}(\Theta | \mathcal{D})  \propto \mathcal{L} (\mathcal{D} | \Theta) \, p(\Theta)}$, where $\mathcal{L} (\mathcal{D} | \Theta)$ is the likelihood and $p(\Theta)$ the prior distribution.
Since in this work we consider the number counts of cosmic voids, the likelihood can be assumed to follow Poisson statistics \citep{sahlen2016}:
\begin{equation} \label{eq:likelihood}
\begin{aligned}
& \mathcal{L} (\mathcal{D} | \Theta) = \\
& = \prod_{i,j} \frac{N(r_i,z_j | \Theta)^{N(r_i,z_j | \mathcal{D})} \, \exp{\left[ -N(r_i,z_j | \Theta)\right]}}{N(r_i,z_j | \mathcal{D})!} \,,
\end{aligned}
\end{equation}
where the product is over the radius and redshift bins, labelled with $i$ and $j$, respectively.  The quantity $N(r_i,z_j | \mathcal{D})$ is the number of voids in the $i^\mathrm{th}$ radius bin and $j^\mathrm{th}$ redshift bin, while $N(r_i,z_j | \Theta)$ corresponds to the expected value in the considered cosmological model, given a set of parameters $\Theta$.

The first cosmological scenario we investigate in this work is the standard flat $\Lambda$CDM model, characterized by six fundamental parameters: the total matter density parameter ($\Omega_{\rm m}$), the dimensionless Hubble parameter ($h \equiv H_0/100 \ \mathrm{km}\,\mathrm{s}^{-1}\mathrm{Mpc}^{-1}$), the baryon density parameter times $h^2$ ($\Omega_{\rm b} \, h^2$), the spectral index of the primordial power spectrum ($n_{\rm s}$), the reionization optical depth ($\tau$) and the primordial power spectrum amplitude ($A_{\rm s}$), which we appropriately reinterpret with the present-time matter power spectrum normalization ($\sigma_8$). In our modeling we require flatness by imposing the cosmological constant density parameter to be $\Omega_{\Lambda}=1-\Omega_{\rm m}$ and define the value of the cold dark matter component as ${\Omega_{\rm cdm}=\Omega_{\rm m}-\Omega_{\rm b}}$. We underline that, analogously to $\sigma_8$, the Universe density parameters defining the cosmological scenario are expressed with their present-day values.

Besides the current standard model of cosmology, we also consider a Universe framework implementing a constant dark energy equation of state, which is commonly denoted $w$CDM. This model represents the simplest dark energy model alternative to $\Lambda$CDM, and is therefore characterized by an additional parameter, $w$. This scenario reduces to standard $\Lambda$CDM when $w=-1$.

In the Bayesian analysis reported in \Cref{sec:results}, both cosmological scenarios will be explored. We focus in particular on the cosmological degeneracies acting on the parameter planes $\Omega_{\rm m}$--$\sigma_8$, for the $\Lambda$CDM model, and $\Omega_{\rm m}$--$w$, for the $w$CDM model.
This selection is motivated by recent promising studies concerning the complementarity of different void statistics with standard cosmological probes, which showed the effectiveness of a joint analysis in constraining the growth of cosmic structures and the dark energy equation of state  \citep{pisani2015,Bayer2021,Kreisch2021,EuclidHamaus2021,Contarini2022,Bonici2022,Pelliciari}.

We sample the posterior distribution of the model parameters with a Markov chain Monte Carlo (MCMC) technique, by considering wide uniform priors for the cosmological parameters of interest, i.e. $\Omega_{\rm m}$ and $\sigma_8$ for the $\Lambda$CDM scenario, $\Omega_{\rm m}$ and $w$ for the alternative scenario. We marginalize over the remaining cosmological parameters ($h$, $\Omega_{\rm b} \, h^2$, $n_{\rm s}$, $\tau$ for the first, and also $\sigma_8$ for the second model) by assigning uniform priors centered on their corresponding values in the PATCHY simulation cosmology \citep{Planck2013} and with a total amplitude equal to ten times the $68\%$ uncertainty provided by the Planck2018 results \citep{Planck2018_results}. We tested the effect of using different priors, considering the constraints given by the Planck2013 cosmology \citep{Planck2013}, finding no or non-significant variations ($<1.5\%$) on both the predicted cosmological values and their relative precision.

Two fundamental quantities to compute at each step of the MCMC procedure are the Hubble parameter, $H(z)$, and the matter power spectrum, $P(k)$. We calculate the former as:
\begin{equation}
H(z)=H_0\left[ \Omega_{\rm m}(1+z)^3 + \Omega_{\rm de} (1+z)^{3 (1+w)} \right]^{1/2}\, ,
\end{equation}
imposing $w=-1$ for the $\Lambda$CDM case,  where ${\Omega_{\rm de} = \Omega_\Lambda}$. The total matter power spectrum is instead evaluated with the approximations for the matter transfer function of \cite{EH1999}. 
All the cosmological calculations (e.g., the 2PCF modeling reported in \ref{appendix:multipoles}), the Bayesian analyses and the data catalog manipulations (including the void cleaning procedure described in \Cref{sec:cleaning}) described in this work are performed in the numerical environment provided by the \texttt{CosmoBolognaLib}\footnote{\url{https://gitlab.com/federicomarulli/CosmoBolognaLib}}\citep{CBL}, a large set of \textit{free software} C++/Python libraries composed of constantly growing and improving codes for cosmological analyses.

\subsection{Model calibration}\label{sec:model_calibration}

As discussed in \Cref{sec:theory}, to model the size function of cosmic voids identified in a biased tracer field, we need to estimate the effective bias of our mass tracers. For this purpose we measure the redshift-space anisotropic 2PCF of the analyzed galaxies by means of the popular \cite{Landy_Szalay1993} estimator. We then model the corresponding multipole moments with the theoretical prescriptions presented in \cite{Taruya2010}, which take into account both the nonlinear gravitational clustering, redshift-space distortions and galaxy bias. We refer the reader to \ref{appendix:multipoles} for a detailed description of the 2PCF modeling performed in this work.

\begin{figure*}
\centering
\includegraphics[width=0.55\textwidth]{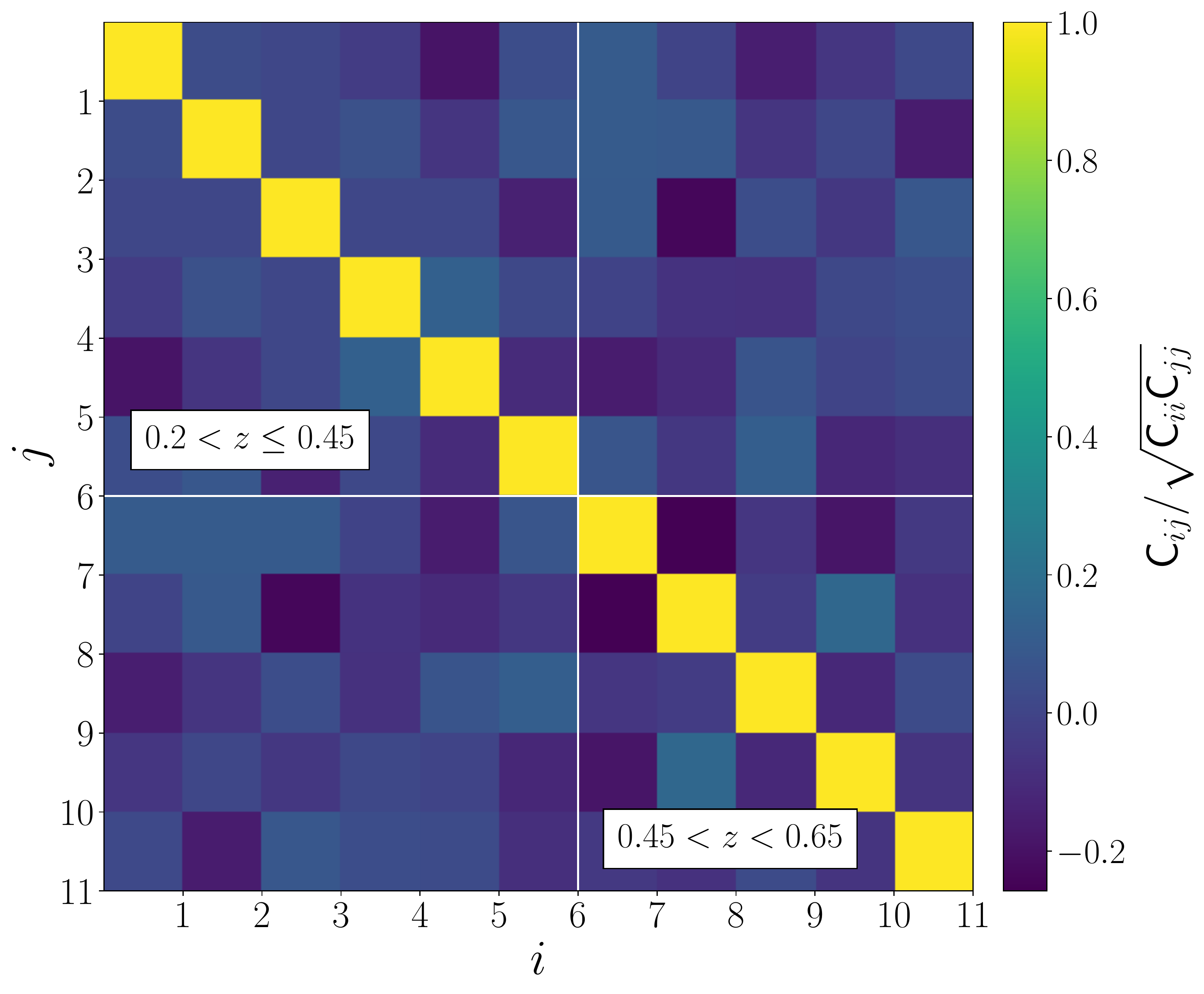}
\includegraphics[width=0.44\textwidth]{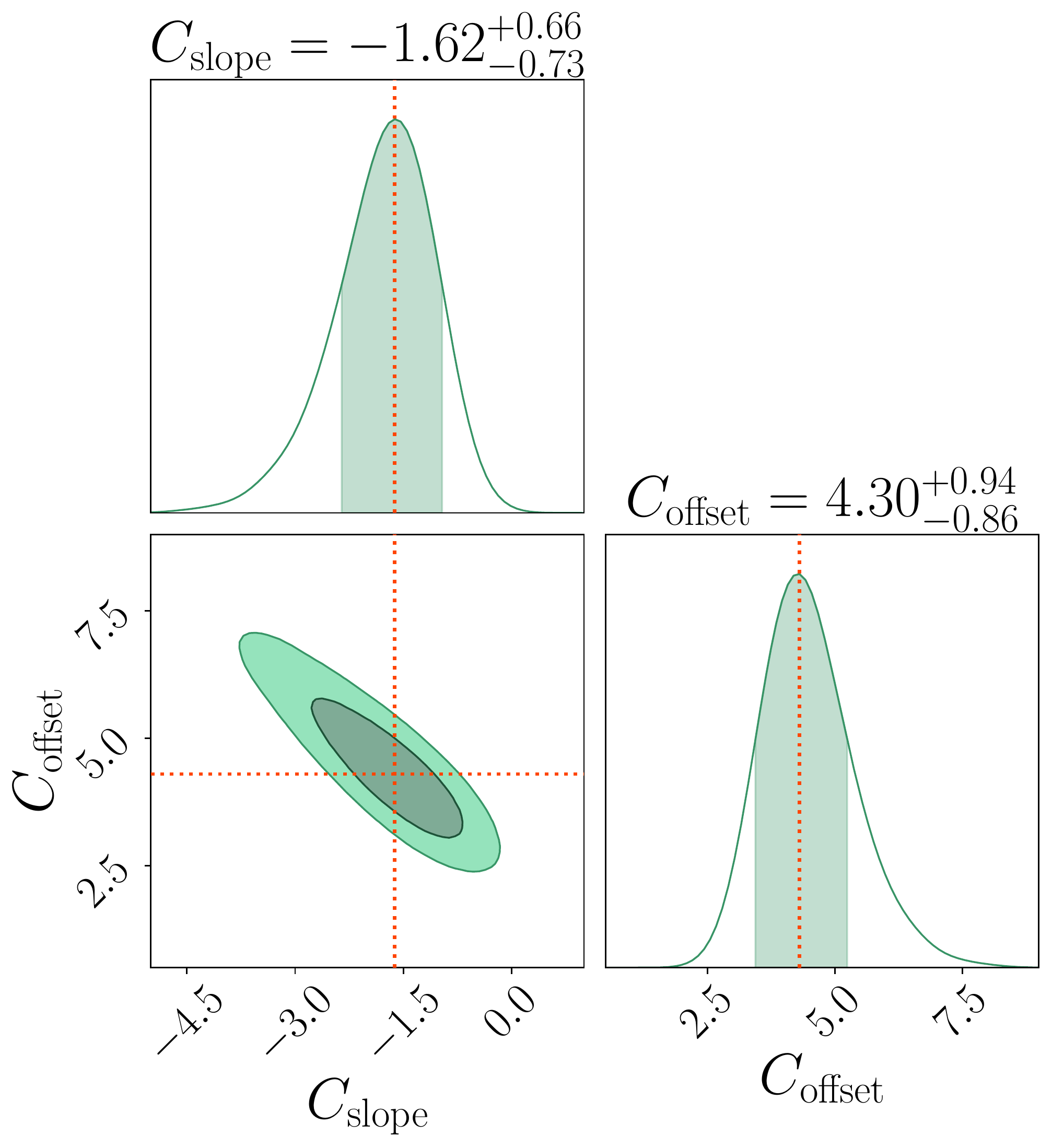}
\caption{\textit{Left}: covariance matrix (normalized by its diagonal components) of MultiDark PATCHY void counts, measured after the application of the cleaning procedure. The different radius bins $(i,j)$ are divided by vertical and horizontal white lines to separate the two redshift intervals covered by the void number counts. \textit{Right}: $68\%$ and $95\%$ 2D confidence contours obtained from the void size function model calibration. The maximum posterior values found for $C_\mathrm{slope}$ and $C_\mathrm{offset}$ are indicated with orange dotted lines and reported at the top of the \textit{upper} and the \textit{right} sub-panels. The latter show the projected 1D posterior distribution of these parameters and the relative $68\%$ uncertainty with a green shaded band. \label{fig:calib}}
\vspace{0.2cm}
\end{figure*}

We apply the galaxy clustering modeling procedure to the $100$ PATCHY mock galaxy catalogs obtaining the correlation matrix of the 2PCF multipoles and the values of $b_\mathrm{eff} \, \sigma_8$ for each realization. 
By computing the median and the standard deviation of the $100$ $\sigma_8$-dependent bias values of the PATCHY galaxy mocks, we attain $b_\mathrm{eff} \, \sigma_8 = 1.33 \pm 0.02$ and $ 1.26 \pm 0.02$, respectively for the two redshift bins considered in this analysis, i.e. $0.2 < z \leq 0.45$ and $0.45 < z < 0.65$ (see \Cref{sec:data_prep}). The corresponding values of the effective bias, achievable in this case knowing the true cosmology of the MultiDark PATCHY simulations, are $b_\mathrm{eff} = 1.94 \pm 0.03$ and $ 2.02 \pm 0.03$.

We use the values of $b_\mathrm{eff} \, \sigma_8$ to calibrate the linear function $\mathcal{F}$ reported in \Cref{eq:new_F} by means of a Bayesian analysis of the number counts of PATCHY cleaned voids. To this end, we measure the median number counts of the $100$ PATCHY samples of voids processed by the cleaning algorithm, fixing the minimum void radius at $2.5$ times the MGS computed at the mean redshift of the galaxies in the low- and high-redshift sub-samples considered. This choice is in agreement with previous works \citep{ronconi2019,contarini2019,Contarini2020,Contarini2022} and leads to the selection of voids with approximately $R_{\rm eff}>35 \ h^{-1} \ \mathrm{Mpc}$.  A detailed discussion of the impact of this choice is presented in \ref{appendix:systematics}.

Our goal is to fit both the median PATCHY void count data sets (i.e. featuring the two selected redshift ranges) to calibrate the free parameters, $C_{\rm slope}$ and $C_{\rm offset}$, of the extended Vdn model described in \Cref{sec:Vdn_extension}.
As for $B_{\rm slope}$ and $B_{\rm offset}$, these coefficients are meant to take into account the galaxy bias variation with redshift, $b_\mathrm{eff}(z)$, so our parametrization is expected to hold in the full range of redshift covered by the survey. Nevertheless we notice that, if the galaxy samples extracted at different redshifts were characterized by incompatible intrinsic properties (e.g. different flux or mass limits) this relation is in fact expected to vary \citep{contarini2019}, eventually deviating from linearity. In our case, the galaxy sample selection is restricted to two redshift bins only, so the linear relation calibrated encompasses any potential variation in the intrinsic properties of the sample.

We recall the reader that the calibrated values, as well as the relative covariance relation for these parameters, are expected to be different from those found in previous papers \citep[i.e.][]{contarini2019,Contarini2020,Contarini2022} for $B_{\rm slope}$ and $B_{\rm offset}$. In these works indeed, $\mathcal{F}$ was parameterized by means of $b_\mathrm{eff}$ only, therefore the first coefficient of the linear function is positive, because of the increasing dependence of $b_\mathrm{eff}$ with redshift\footnote{This is true except for galaxy samples characterized by strong survey selection effects.}. Conversely, the slope of the linear function $\mathcal{F}$ is expected to be negative in our analysis given its dependence on $b_\mathrm{eff} \, \sigma_8$, which is instead decreasing with redshift for the selected sample of galaxies.

We perform the Bayesian analysis starting by assuming the likelihood form expressed in \Cref{eq:likelihood} and computing the covariance between our $100$ PATCHY number count measures of the cleaned voids samples. We report the void counts covariance matrix, $\textsf{C}_{ij}$, normalized by its diagonal component, $\sqrt{\textsf{C}_{ii}\textsf{C}_{jj}}$, in the left panel of \Cref{fig:calib}. We separate in this plot the low- and the high-redshift ranges considered in our analysis, characterized by six and five logarithmic radius bins, respectively. We notice that the covariance between the void radii at the same redshift and between the two redshift bins exhibits negligible off-diagonal terms \citep[as suggested by][]{Kreisch2021}. We tested the effect of modeling the void counts with a Gaussian likelihood including the full covariance matrix presented here and we found no statistically relevant change in the results, confirming the low impact of the off-diagonal terms in this analysis. We finally note that the super-sample covariance effects are expected to be small on the void size function, so its covariance can be reliably approximated without considering the presence of super-survey modes \citep{Bayer2022}.

To model the average void number counts extracted from the PATCHY mock catalogs and to calibrate the void size function nuisance parameters, we applied the extended Vdn model to the low and high redshift bins analyzed in this work. We consider the median redshifts of the two galaxy samples, assigning to each the corresponding value of the $\sigma_8$-dependent bias as a Gaussian prior, with mean and standard deviation given by the 2PCF analysis. Then we set wide uniform priors for $C_{\rm slope}$ and $C_{\rm offset}$, and sample the posterior distribution of these parameters, assuming the fiducial cosmology of the MultiDark PATCHY simulations.  

We show the results of the nuisance parameter calibration in the right panel of \Cref{fig:calib}, where we report the $68\%$ and $95\%$ confidence contours on the parameters $C_{\rm slope}$ and $C_{\rm offset}$. The values of the maximum posterior probability are reported at the top of the two panels representing the associated 1D posterior distribution projections. We notice that $C_{\rm slope}$ and $C_{\rm offset}$ are strongly anti-correlated, and that their posterior distribution is asymmetric. In particular, the skewness of the distribution is due to the physical requirement of $\delta_{\rm v, DM}^{\rm NL} > -1$ for the underdensity threshold, resulting from the Vdn model reparametrization.

\begin{figure*}
\centering
\includegraphics[width=0.9\textwidth]{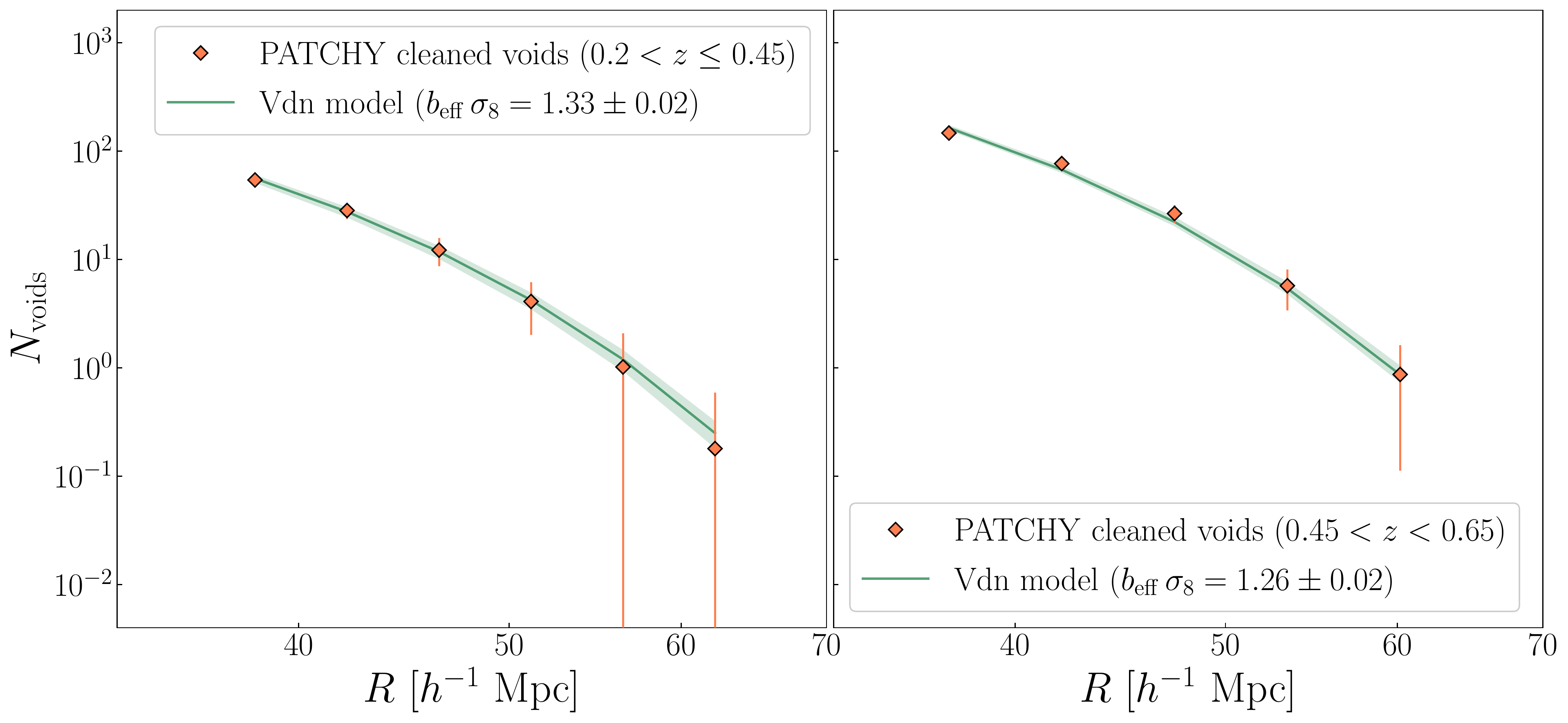}
\caption{Median void counts as a function of their cleaned radius, measured from $100$ mocks of the MultiDark PATCHY simulation and for the two redshift ranges considered this analysis: $0.2 < z \leq 0.45$ (\textit{left}) and $0.45 < z < 0.65$ (\textit{right}). For each panel, we represent the median void number counts as orange diamonds, with error bars given by their standard deviation across the different PATCHY realizations. We indicate with a green line the calibrated void size function model, dependent on the value of $b_\mathrm{eff} \, \sigma_8$ of the redshift-selected sample of galaxies. The green shaded area represents the $68\%$ confidence region around the calibrated model, determined by the confidence contour reported in the right panel of \Cref{fig:calib}. \label{fig:PATCHY_VSF}}
\vspace{0.2cm}
\end{figure*}

In \Cref{fig:PATCHY_VSF} we report the calibrated Vdn model derived from the fitting procedure, together with the average void counts measured from the PATCHY mock catalogs. We show in the two panels of this figure the results for the low and high redshift bins analyzed, $0.2 < z \leq 0.45$ and $0.45 < z < 0.65$. The markers indicate the number counts of cleaned voids, while the error bars are computed as the standard deviation of the $100$ number count measures in each radius bin. We highlight that this uncertainty on the void counts is similar, although larger than the Poissonian error (i.e. the mean square root of the void counts), which represents an underestimation of the true error associated with the counts.
The line indicates the theoretical prediction from the extended Vdn model, dependent on the value of $b_\mathrm{eff} \, \sigma_8$, characterizing the galaxies belonging to the considered redshift sub-sample. We report with a shaded band the $68\%$ uncertainty on the void size function model, associated with the constraints on the parameters $C_{\rm slope}$ and $C_{\rm offset}$, derived from the calibration on the MultiDark PATCHY simulations.

The agreement between the measured void number counts and the theoretical predictions of the calibrated Vdn model proves the effectiveness of our methodology, which we exploit in this work following two different approaches. The first is to assume the exact values, i.e. without any associated uncertainty, for the parameters $C_{\rm slope}$ and $C_{\rm offset}$ obtained from the calibration of the model on the PATCHY mocks. This case corresponds to the ideal assumption of using volume-unlimited cosmological simulations perfectly reproducing the survey data that one wants to study or, alternatively, to have a size function model including a fully theoretical description of both the effects of redshift-space distortions and mass tracer bias. This approach has been explored from a theoretical point of view in recent works \citep[e.g.,][]{Verza2022}, but a complete and calibration-independent model has not yet been provided. Although optimistic, this first approach shows the full potential of the methodology and is useful to understand the limitations related to the calibration procedure; the analysis performed with fixed values of $C_{\rm slope}$ and $C_{\rm offset}$ will be labeled hereafter as \textit{fixed calibration}.

The second approach consists of including the uncertainties deriving from the calibration procedure: the entire posterior distribution of the parameters $C_{\rm slope}$ and $C_{\rm offset}$, sampled using the PATCHY void number counts and reported in the right panel of \Cref{fig:calib}, is assumed as prior for the Bayesian analysis of the BOSS void size function.
The analysis conducted with this methodology will be denoted as \textit{relaxed calibration}, and provides more conservative cosmological constraints, due to the limits of the calibration procedure performed with the MultiDark PATCHY simulations. We refer the reader to \ref{appendix:systematics} for further insights about the impact of the calibration parameter priors on the cosmological constraints from the void size function.

To validate the proposed modeling approaches, we derive mock cosmological constraints from the PATCHY void size function, with the aim of recovering the true cosmology of the simulation and evaluating the constraining power of the measured void number counts. We report the results of this test in \ref{appendix:PATCHY_constraints}, where we prove the effectiveness of the model in retrieving the MultiDark PATCHY simulation input cosmological parameters within $68\%$ errors, for both the assumed cosmological models (i.e. $\Lambda$CDM and $w$CDM). We also refer the interested reader to \ref{appendix:PATCHY_constraints} for a discussion about the dependency of the void size function constraining power on the main characteristics of the analyzed data: the number density and volume coverage of the galaxy sample. As these two quantities differ between the PATCHY and BOSS catalogs, the precision of the cosmological constraints obtained from the corresponding void samples is also slightly different.

\begin{figure*}
\centering
\includegraphics[width=0.9\textwidth]{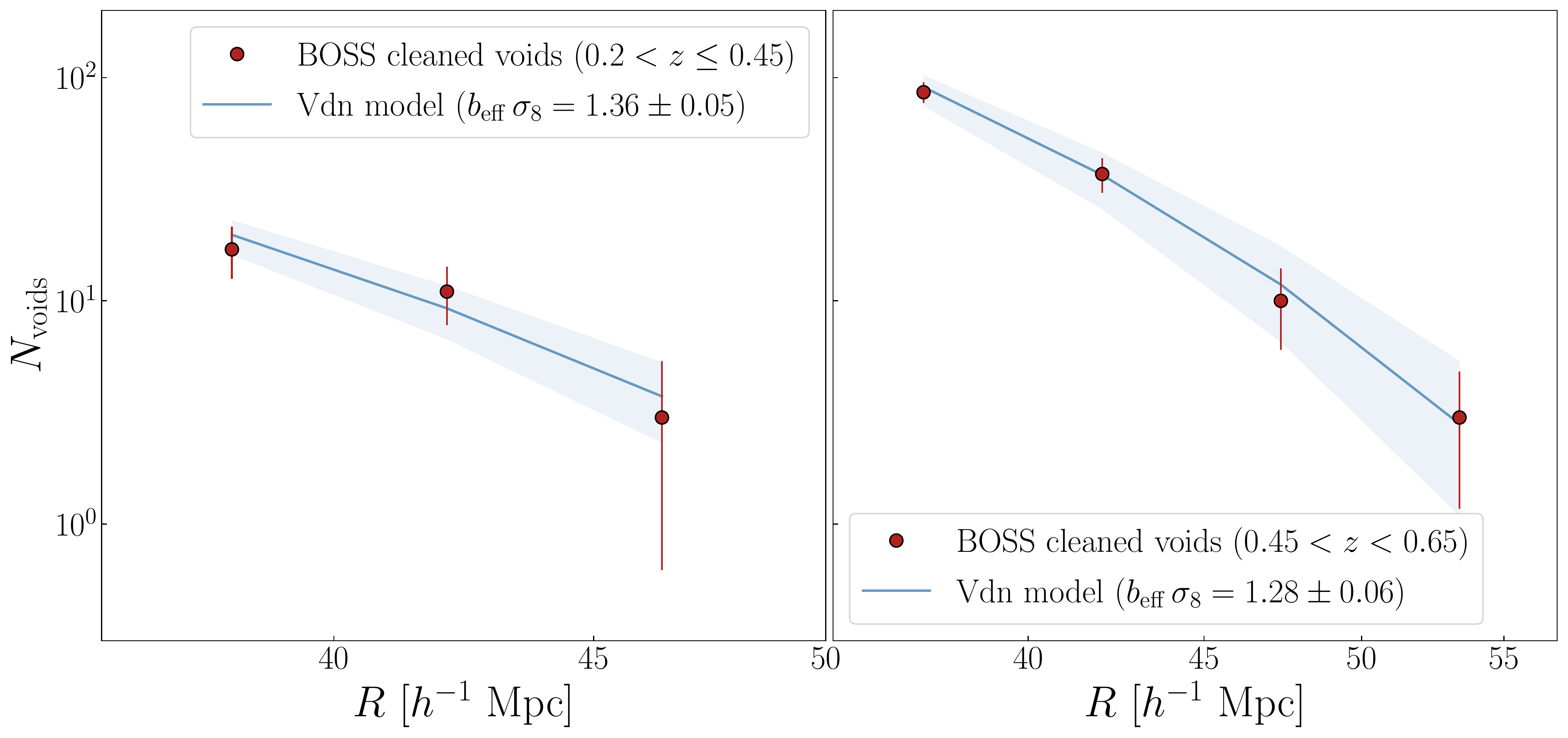}
\caption{Void counts as a function of their cleaned radius measured in the BOSS DR12 galaxy distribution, for the two redshift sub-samples considered in this analysis: $0.2 < z \leq 0.45$ (\textit{left}) and $0.45 < z < 0.65$ (\textit{right}). For each panel, we report with red circles the measured void size function, with error bars computed by rescaling those associated with the corresponding PATCHY void counts. We show in blue the void size function model calibrated with the PATCHY mock catalogs and dependent on the value of $b_\mathrm{eff} \, \sigma_8$ of the redshift-selected sample of galaxies. 
In this case, the best-fit model (solid line) and its $68\%$ confidence region (shaded) are computed by fitting the measured void counts with a $\Lambda$CDM cosmology and leaving the parameters $\sigma_8$ and $\Omega_\mathrm{m}$ free (see the \textit{relaxed calibration} case of \Cref{sec:model_calibration}). \label{fig:BOSS_VSF}}
\vspace{0.2cm}
\end{figure*}

\section{Results} \label{sec:results}

\subsection{BOSS DR12 void counts} \label{sec:BOSS_results}

Now that the void size function model has been calibrated, we are ready for the cosmological exploitation of the void number counts measured from the BOSS galaxy catalogs. To prepare the void sample, we follow the same pipeline applied to the PATCHY mocks. Firstly, we clean the void catalogs extracted with \texttt{VIDE}, assuming the galaxy number density distribution reported in the left panel of \Cref{fig:VIDE_VSF} and approximated with a spline function, which is represented as a dashed line. During the cleaning operation, we consider the same values for the density threshold, $\delta_\mathrm{v,tr}^\mathrm{NL}=-0.7$, and for the void size selection, $R>30 \ h^{-1} \ \mathrm{Mpc}$. Then we extract from this sample the number of voids as a function of their radius for the two redshift ranges considered in this analysis (see \Cref{sec:data_prep}). We finally compute the uncertainties on the binned void size function measures through a volume-based rescaling of the errors estimated for the PATCHY void number counts, as explained in \Cref{sec:data_prep} to describe the right panel of \Cref{fig:VIDE_VSF}.

These data sets, obtained with the fiducial Planck2013 cosmology \citep{Planck2013}, need to be compared with the corresponding theoretical model of the void size function evaluated with the calibrated parameters $C_\mathrm{slope}$ and $C_\mathrm{offset}$. To do this, we measure the value of the $\sigma_8$-dependent bias, considering the BOSS galaxies belonging to the two selected redshift sub-samples, and the 2PCF covariance matrix built by means of PATCHY mocks (see \Cref{sec:model_calibration} and \ref{appendix:PATCHY_constraints}): we find $b_\mathrm{eff} \, \sigma_8 = 1.36 \pm 0.05$ for the interval $0.2 < z \leq 0.45$ and $b_\mathrm{eff} \, \sigma_8 = 1.28 \pm 0.06$ for $0.45 < z < 0.65$. Thereafter we use these values to compute the theoretical void size function by means of the extended Vdn model calibrated in \Cref{sec:model_calibration}, and we rely on the effective volume covered by the void finder applied to the BOSS galaxies to convert the predicted void number density into number counts, directly comparable with our observed data sets.

At this point we perform a Bayesian analysis of the measured void number counts by assuming the likelihood function reported in \Cref{eq:likelihood} and testing the two cosmological models outlined in \Cref{sec:cosmological_modeling}. For both the considered cosmological scenarios, i.e. $\Lambda$CDM and $w$CDM, we follow the two approaches presented in \Cref{sec:model_calibration}, which depend on the constraints assumed from the model calibration with PATCHY mock catalogs, i.e. the \textit{fixed calibration} and the \textit{relaxed calibration} methodologies.

In \Cref{fig:BOSS_VSF} we show a comparison between the measured size function of voids identified in the distribution of BOSS DR12 galaxies and the best-fit model computed with the MCMC sampling technique. We report the results for the $\Lambda$CDM cosmological model, obtained treating the parameters $C_{\rm slope}$ and $C_{\rm offset}$ of the extended Vdn model as free, but restrained by a prior distribution determined from the PATCHY calibration constraints (i.e. the \textit{relaxed calibration} case). We recall that in this specific analysis the free cosmological parameters of the model are $\Omega_{\rm m}$ and $\sigma_8$, while the remaining ones are bound to Planck2018 constraints (see \Cref{sec:cosmological_modeling} for a more detailed description). The MCMC technique used to sample the posterior distribution of all these parameters provides us with the best-fit value and the corresponding uncertainty for the void size function model, represented in \Cref{fig:BOSS_VSF} as a solid line surrounded by a shaded band covering the $68\%$ confidence region.

We point out the good agreement between the measured void number counts and the extended Vdn model in \Cref{fig:BOSS_VSF}, for both the low (left panel) and high (right panel) redshift ranges analyzed, featuring three and four independent radius bins, respectively. 
We highlight that the void number count binning applied here is the same as the one used for the PATCHY analysis. In this case, however, the void counts do not derive from an averaging procedure over different realizations and therefore cannot be lower than unity. This fact, combined with the statistical rarity of large voids, leads to more restricted void size ranges with respect to those presented in \Cref{fig:PATCHY_VSF}.
Moreover, analogously to what we found for the PATCHY mocks, voids are more abundant in the high redshift sub-sample, given the greater volume covered by the galaxies in the light-cone.

The data-model agreement obtained for the remaining modeling approach (i.e. the \textit{fixed calibration} case) and for the alternative cosmological scenario considered (i.e. $w$CDM, assuming both types of calibration parameter priors) are similar to the results presented, which is why we do not explicitly show them here. Nevertheless, the cosmological constraints corresponding to each case will be described in the following section.

\subsection{Cosmological constraints}\label{sec:cosmological_constraints}

\begin{figure*}
\centering
\hspace{0.5cm}
\includegraphics[width=0.47\textwidth]{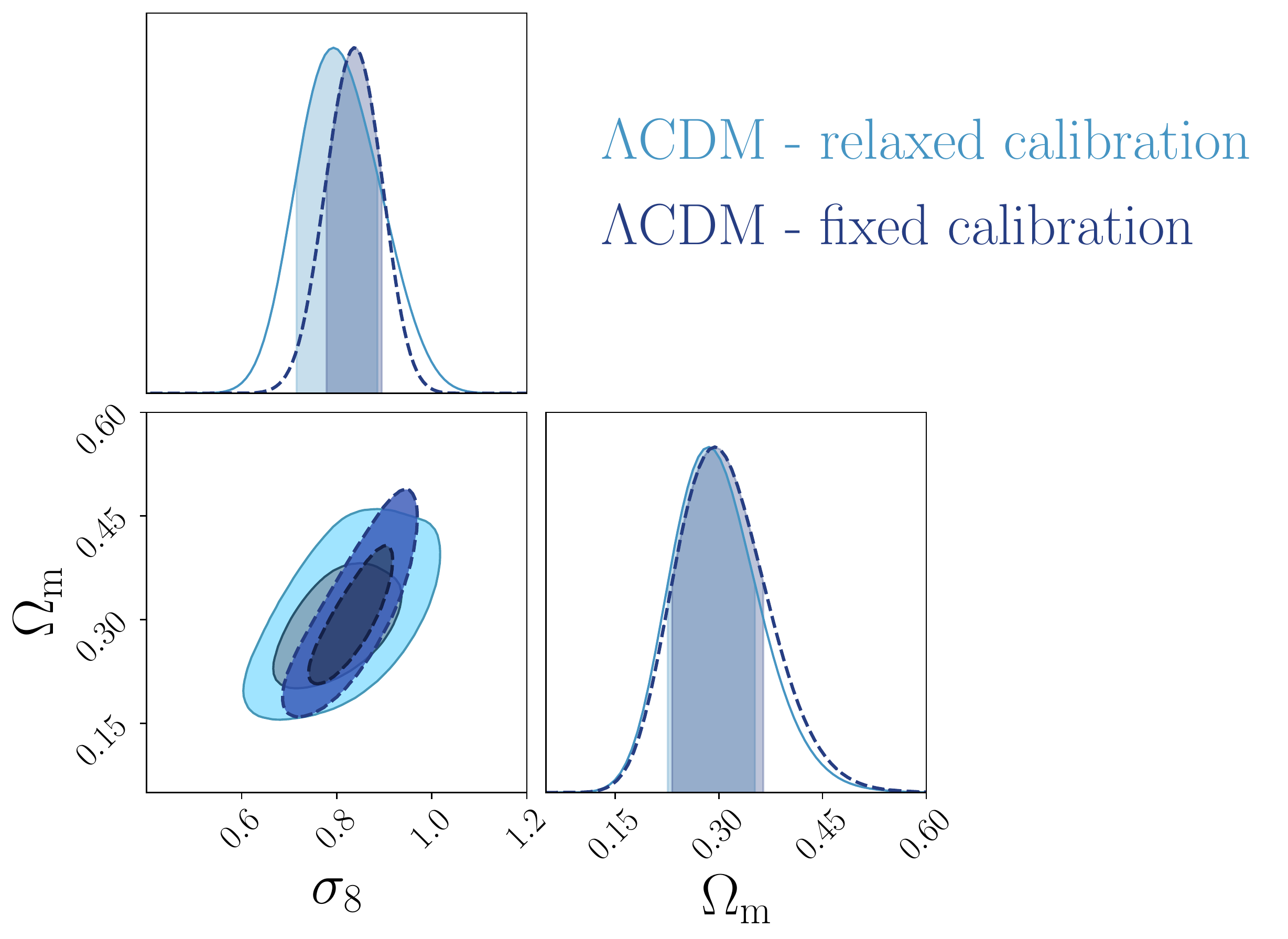}
\includegraphics[width=0.47\textwidth]{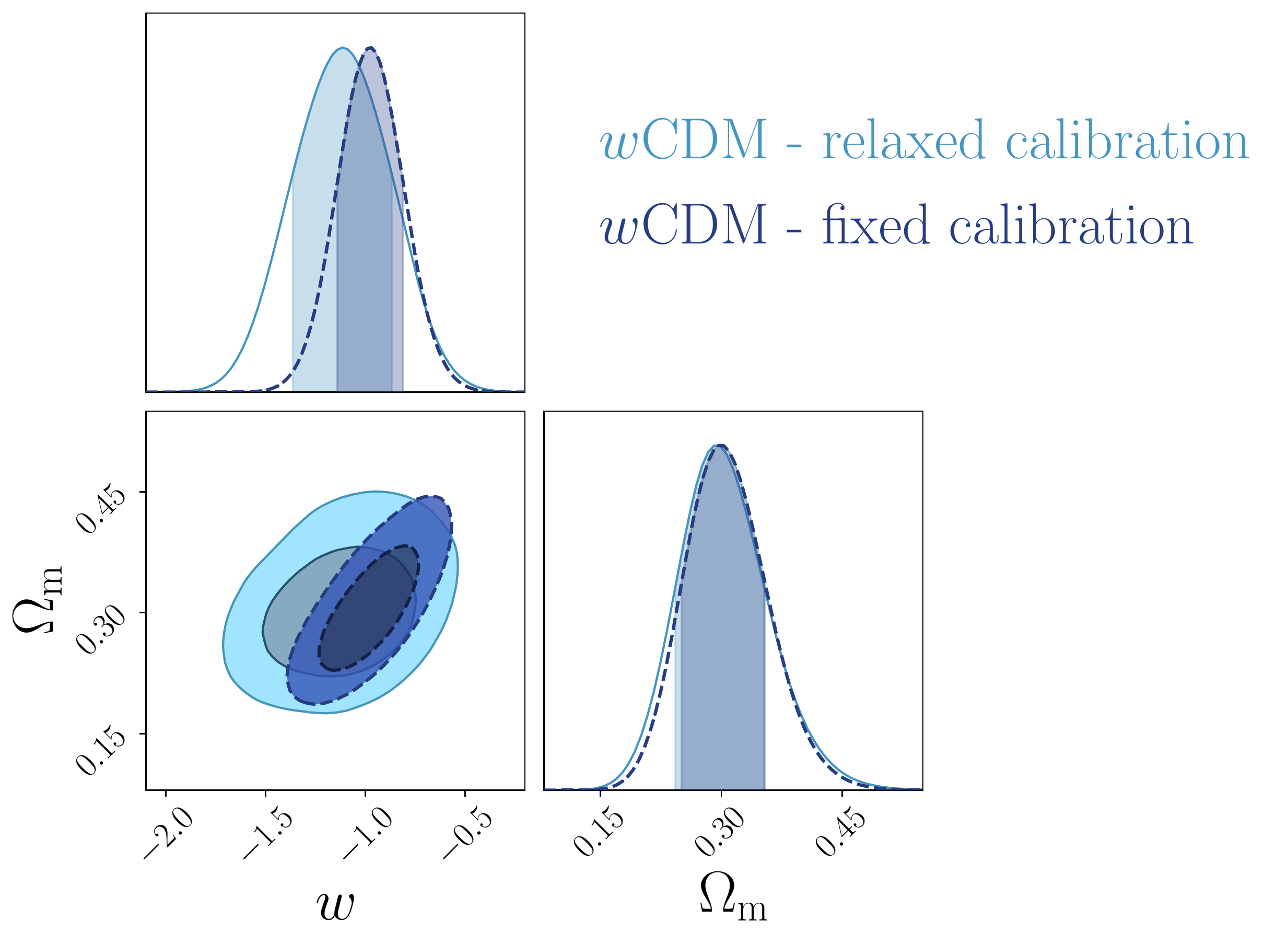}
\caption{Constraints from the size function of voids identified in the BOSS DR12 data set, for $\Lambda$CDM (\textit{left}) and $w$CDM (\textit{right}) flat cosmologies. For both panels, the 2D contours indicate the $68\%$ and $95\%$ confidence levels on the posterior distribution of the main model parameters, while the 1D distributions show the marginalized posteriors of each parameter with the $68\%$ confidence interval indicated as a shaded band. In dark blue we report the results obtained by fixing the parameters $C_\mathrm{slope}$ and $C_\mathrm{offset}$ to their maximum posterior values obtained from the model calibration with PATCHY mocks (\textit{fixed calibration}). In light blue we show the constraints when assuming an uncertainty on $C_\mathrm{slope}$ and $C_\mathrm{offset}$, given by their posterior distribution reported on the right panel of \Cref{fig:calib} (\textit{relaxed calibration}). 
}\label{fig:BOSS_constraints}
\vspace{0.1cm}
\end{figure*}

\Cref{fig:BOSS_constraints} shows the constraints resulting from both void size function calibration approaches presented in this work. In the left panel we show the results for the $\Lambda$CDM scenario, while in the right those for the alternative $w$CDM model. We represent in different colors the $68\%$ and $95\%$ confidence contours for the two cases labelled \textit{fixed} and \textit{relaxed calibration}. We note that when relaxing the calibration constraints, the corresponding confidence contours slightly enlarge as a consequence of the contribution given by the variation of the parameters $C_\mathrm{slope}$ and $C_\mathrm{offset}$.
The main effect of the \textit{relaxed calibration} approach is to loosen the constraining power on the parameters $\sigma_8$ and $w$, for the $\Lambda$CDM and $w$CDM scenarios, respectively. On the other hand, as visible from the 1D marginalized posterior distributions reported on the sub-panels of \Cref{fig:BOSS_constraints}, the corresponding impact on $\Omega_{\rm m}$ is statistically negligible.

One of the most important features to notice is the orientation of the $\Omega_{\rm m}$--$\sigma_8$ confidence contour reported in the left panel of \Cref{fig:BOSS_constraints}. First, this degeneracy direction is in agreement with the results of other works that analyzed the void size function in simulated catalogs \citep{Bayer2021,Kreisch2021}. Second, as demonstrated by \cite{Contarini2022} and  \cite{Pelliciari}, it is orthogonal to that of other standard cosmological probes, such as galaxy clustering, weak lensing and cluster number counts. The latter is a fundamental characteristic to be exploited in order to maximize the constraining power given by probe combinations, thanks also to the negligible correlation of the void size function with standard statistics of overdense objects. We provide some insights on the physical origin of the orthogonality of these constraints in \ref{appendix:orthogonality}.

\begin{table*}
    \centering
    \caption{Maximum posterior values and $68\%$ uncertainties associated with the confidence contours reported in \Cref{fig:BOSS_constraints}.
    We present the results for the two cosmological scenarios analyzed in this work, i.e. $\Lambda$CDM and $w$CDM, in two separate sets of columns, grouping together the freely varied parameters of the corresponding model.
    The first row of the table presents the results obtained by modeling the size function of BOSS DR12 voids using the \textit{relaxed calibration} method, while the second row reports the constraints when adopting the \textit{fixed calibration} approach (see \Cref{sec:model_calibration} for further details).} \label{tab:BOSS_results}
    \centering
    \begin{tabular}{ccccccccccc}
        \toprule
        \noalign{\vspace{0.1cm}}
		\multirow{2}{*}{\shortstack{ \\ \\ Model \\ \\ calibration}} & & \multicolumn{4}{c}{$\Lambda$CDM} & & \multicolumn{4}{c}{$w$CDM} \\
		\noalign{\vspace{0.1cm}}
		\cline{3-6}
		\cline{8-11}
		\noalign{\vspace{0.1cm}}
		& & $\Omega_\mathrm{m}$ & $\sigma_8$ & $C_\mathrm{slope}$ & $C_\mathrm{offset}$ & & $\Omega_\mathrm{m}$ & $w$ & $C_\mathrm{slope}$ & $C_\mathrm{offset}$\\ 
		\noalign{\vspace{0.1cm}}
		\hline
		\noalign{\vspace{0.1cm}}
		Relaxed & & $0.285^{+0.063}_{-0.057}$ & $0.793^{+0.090}_{-0.076}$ & $-2.02^{+0.66}_{-0.61}$ & $4.74^{+0.87}_{-0.82}$ & & $0.293^{+0.055}_{-0.047}$ & $-1.11\pm 0.24$ & $-2.07^{+0.70}_{-0.73}$ & $4.90^{+1.00}_{-0.97}$ \\
		\noalign{\vspace{0.1cm}}
		Fixed & & $0.293^{+0.067}_{-0.059}$ & $0.839^{+0.055}_{-0.058}$ & $-1.62$ & $4.30$ & & $0.299^{+0.051}_{-0.046}$ & $-0.97^{+0.15}_{-0.16}$ & $-1.62$ & $4.30$ \\ 
		\noalign{\vspace{0.1cm}}
		\hline
		\bottomrule
    \end{tabular}
\vspace{1cm}
\end{table*}

We report the maximum posterior values and the relative $68\%$ errors derived from the BOSS void size function analysis in \Cref{tab:BOSS_results}. As we discuss in more detail in \cite{Contarini2022_tensions}, the constraints we derived are compatible with the Planck2018 results \citep{Planck2018_results}. Assuming the flat $\Lambda$CDM framework, we obtain a value of $\Omega_{\rm m} = 0.285$ ($\Omega_{\rm m} = 0.293$) for the \textit{relaxed} (\textit{fixed}) calibration approach, with a corresponding precision of about $21\%$. In the same model, we have $\sigma_8=0.793$ ($\sigma_8=0.839$) for the \textit{relaxed} (\textit{fixed}) calibration approach, with a precision of $10\%$ ($7\%$). 

As already mentioned, we expect a significant improvement of the precision on these parameters through the joint analysis of void counts with different cosmological probes. This is investigated in a companion paper \citep{Contarini2022_tensions}, where we present a more specific and detailed analysis of the contribution given by cosmic void counts on two of the main current issues in cosmology: the tension between late and early probes on the Hubble constant $H_0$, and the matter clustering strength parametrised by ${S_8 \equiv \sigma_8\sqrt{\Omega_{\rm m}/0.3}}$ \citep[see e.g.][for a review]{DiValentino2021a,DiValentino2021b}.

For what concerns the flat $w$CDM scenario, we obtain $\Omega_{\rm m}=0.293$ ($\Omega_{\rm m}=0.299$), with a corresponding precision of $17\%$ ($16\%$) for the \textit{relaxed} (\textit{fixed}) calibration strategy. Then we compute ${w=-1.11}$ ($w=-0.97$) for the \textit{relaxed} (\textit{fixed}) calibration approach, with a precision of $22\%$ ($16\%$). For both cases the results are compatible with the constant $w=-1$, so in agreement with the benchmark $\Lambda$CDM model.

\section{Conclusions} \label{sec:conclusions}

The goal of this paper was to provide the first cosmological constraints from the void size function. The void catalog has been extracted from the BOSS DR12 galaxies \citep{BOSSDR12}, which still represents one of the best-suited spectroscopic catalogs for the purpose of void identification.

In this analysis we employ an extension of the popular Vdn model \citep{jennings2013}, which takes into account the effect of the galaxy effective bias on the traced void density profile, as well as geometric and dynamic distortions on the observed void shape (see \Cref{sec:Vdn_extension}). This model includes two nuisance parameters, $C_{\rm slope}$ and $C_{\rm offset}$, which describe the rescaling of the Vdn underdensity threshold by means of the galaxy bias coupled with the matter power spectrum normalization, i.e. the quantity $b_\mathrm{eff} \, \sigma_8$. In order to calibrate these nuisance parameters, we exploit the MultiDark PATCHY simulations, which are designed to reproduce the BOSS data clustering properties \citep{Kitaura2016}.

We prepared the void catalogs by applying a cleaning procedure and conservative selection cuts (see \Cref{sec:cleaning}) to ensure the void samples to be in agreement with the prescriptions of the void size function theory. We then used $100$ PATCHY mocks to calibrate the parameters $C_{\rm slope}$ and $C_{\rm offset}$ and validate the statistical method. The latter was achieved by comparing the theoretical predictions of the extended Vdn model with the void average number counts measured in the MultiDark PATCHY simulation (see \Cref{sec:model_calibration} and \ref{appendix:PATCHY_constraints}). We performed a Bayesian analysis of the BOSS DR12 void number counts assuming a Poissonian likelihood, and investigated two cosmological scenarios (see \Cref{sec:cosmological_modeling}). 

Considering a flat $\Lambda$CDM model, we constrained the present-day values of $\Omega_{\rm m}$ and $\sigma_8$, obtaining ${\Omega_{\rm m}=0.29 \pm 0.06}$ and $\sigma_8=0.79^{+0.09}_{-0.08}$ with the conservative \textit{relaxed} calibration approach (see \Cref{sec:model_calibration}). On the other hand, in a flat $w$CDM model we constrained the parameters $\Omega_{\rm m}$ and $w$, obtaining $\Omega_{\rm m}=0.29^{+0.06}_{-0.05}$ and $w=-1.1 \pm 0.2$ following the most conservative approach. Our estimate of the parameter $w$ is compatible with a value of $-1$ within the $68\%$ confidence region of its posterior, so it reveals no statistical evidence against the standard $\Lambda$CDM model. 

Despite the numerous tests performed in this work to validate the robustness of the methodology, further analyses might be required to fully evaluate all the systematic uncertainties possibly associated with our results. Specifically, we emphasize the importance of investigating the potential influence of the halo population algorithm parameters on the resulting void catalog, as well as exploring the dependence of the model calibration parameters on the fiducial cosmology. Ultimately, these effects should be included in the void size function analysis, by marginalizing over the explored parameters. The full investigation of the possible sources of systematic errors is beyond the scope of this paper and future works will be specifically dedicated to this task.

In conclusion, we stress the fact that the cosmological constraints derived with the void size function are very promising, especially in the perspective of a combination with other cosmological probes \citep{Bayer2021,Kreisch2021}. This is in particular due to the strong orthogonality on the $\Omega_{\rm m}$--$\sigma_8$ parameter plane between the confidence contours obtained with the void size function and overdensity-based probes, as demonstrated in \cite{Contarini2022} and \cite{Pelliciari}.

We explore in detail the synergy with other cosmological probes in a companion paper \citep{Contarini2022_tensions}, using the same void data set built from the BOSS DR12.
Finally, we note that a considerable improvement on the precision of the cosmological constraints achieved in this work will be obtained by applying our methodology to the data of the upcoming wide-field surveys, like \textit{Euclid} \citep{Laureijs2011, Amendola2018}, NGRST \citep{WFIRST2012} and LSST \citep{LSST2012}.

\begin{acknowledgements}
We sincerely thank the Referee for their insightful comments, which undoubtedly have led to the improvement of this work. SC thanks Michele Moresco, Alfonso Veropalumbo, Giorgio Lesci and Nicola Borghi for the contribution and suggestions they provided. The authors thank Scott Dodelson, Steffen Hagstotz, Barbara Sartoris, Nico Schuster, and Giovanni Verza for useful conversations. We acknowledge the grant ASI n.2018-23-HH.0. SC, FM, LM and MB acknowledge the use of computational resources from the parallel computing cluster of the Open Physics Hub (\url{https://site.unibo.it/openphysicshub/en}) at the Physics and Astronomy Department in Bologna.
AP is supported by NASA ROSES grant 12-EUCLID12-0004, and NASA grant 15-WFIRST15-0008 to the Nancy Grace Roman Space Telescope Science Investigation Team ``cosmology with the High Latitude Survey''. AP acknowledges support from the Simons Foundation to the Center for Computational Astrophysics at the Flatiron Institute. NH is supported by the Excellence Cluster ORIGINS, which is funded by the Deutsche Forschungsgemeinschaft (DFG, German Research Foundation) under Germany's Excellence Strategy -- EXC-2094 -- 390783311.
LM acknowledges support from PRIN MIUR 2017 WSCC32 ``Zooming into dark matter and proto-galaxies with massive lensing clusters''. MB acknowledges support by the project ``Combining Cosmic Microwave Background and Large Scale Structure data: an Integrated Approach for Addressing Fundamental Questions in Cosmology", funded by the MIUR Progetti di Ricerca di Rilevante Interesse Nazionale (PRIN) Bando 2017 - grant 2017YJYZAH.
We acknowledge  use  of  the  Python  libraries \texttt{NumPy} \citep{numpy}, \texttt{Matplotlib} \citep{Matplotlib} and \texttt{ChainConsumer} \citep{ChainConsumer}.  
\end{acknowledgements}

\appendix

\section{Multipoles modeling} \label{appendix:multipoles}

In this appendix we present the analysis of the galaxy two-point correlation function multipoles, performed on both the MultiDark PATCHY and BOSS DR12 data, to derive the linear effective bias. This quantity is fundamental to parameterize the void size function model we introduced in \Cref{sec:Vdn_extension}, and consequently to rescale the void radii according to the observed void density profile. 

We measure the 2D redshift-space 2PCF with the \cite{Landy_Szalay1993} estimator:
\begin{equation} \label{2PCF_estimator}
\hat{\xi}(s,\mu)=\frac{N_{\rm RR}}{N_{\rm GG}}\frac{{\rm GG}(s, \mu)}{{\rm RR}(s, \mu)}-2\frac{N_{\rm RR}}{N_{\rm GR}}\frac{{\rm GR}(s,\mu)}{{\rm RR}(s,\mu)}+1 \, ,
\end{equation}
where $\mu$ is the cosine of the angle between the line-of-sight and the observed separation vector $\mathbf{s}$, ${\rm GG}(s, \mu)$, ${\rm RR}(s, \mu)$, and ${\rm GR}(s, \mu)$ are the numbers of galaxy-galaxy, random-random, and galaxy-random pairs in bins of $s$ and $\mu$, $N_{\rm G}$ and $N_{\rm R}$ are the total numbers of galaxies and random objects, and $N_{\rm GG} = N_\mathrm{G}(N_\mathrm{G}-1)/2$, $N_{\rm RR} = N_\mathrm{R}(N_\mathrm{R}-1)/2$, and $N_{\rm GR}$ = $N_\mathrm{G} N_\mathrm{R}$ are the total numbers of galaxy-galaxy, random-random, and galaxy-random pairs, respectively. 
The 2PCF is measured on scales from $10$ to $60 \ h^{-1} \ \mathrm{Mpc}$ and in the two redshift ranges considered in this paper, i.e. $0.2 < z \leq 0.45$ and $0.45 < z < 0.65$, assuming the fiducial cosmology of PATCHY simulations (see \Cref{sec:data_prep}). For both real and mock galaxy samples, we make use of the random catalogs provided by the BOSS collaboration, which are built with the same geometry and redshift distribution of the target data set, but $50$ times denser.  

\begin{figure*}[h]
\centering
\includegraphics[width=0.9\textwidth]{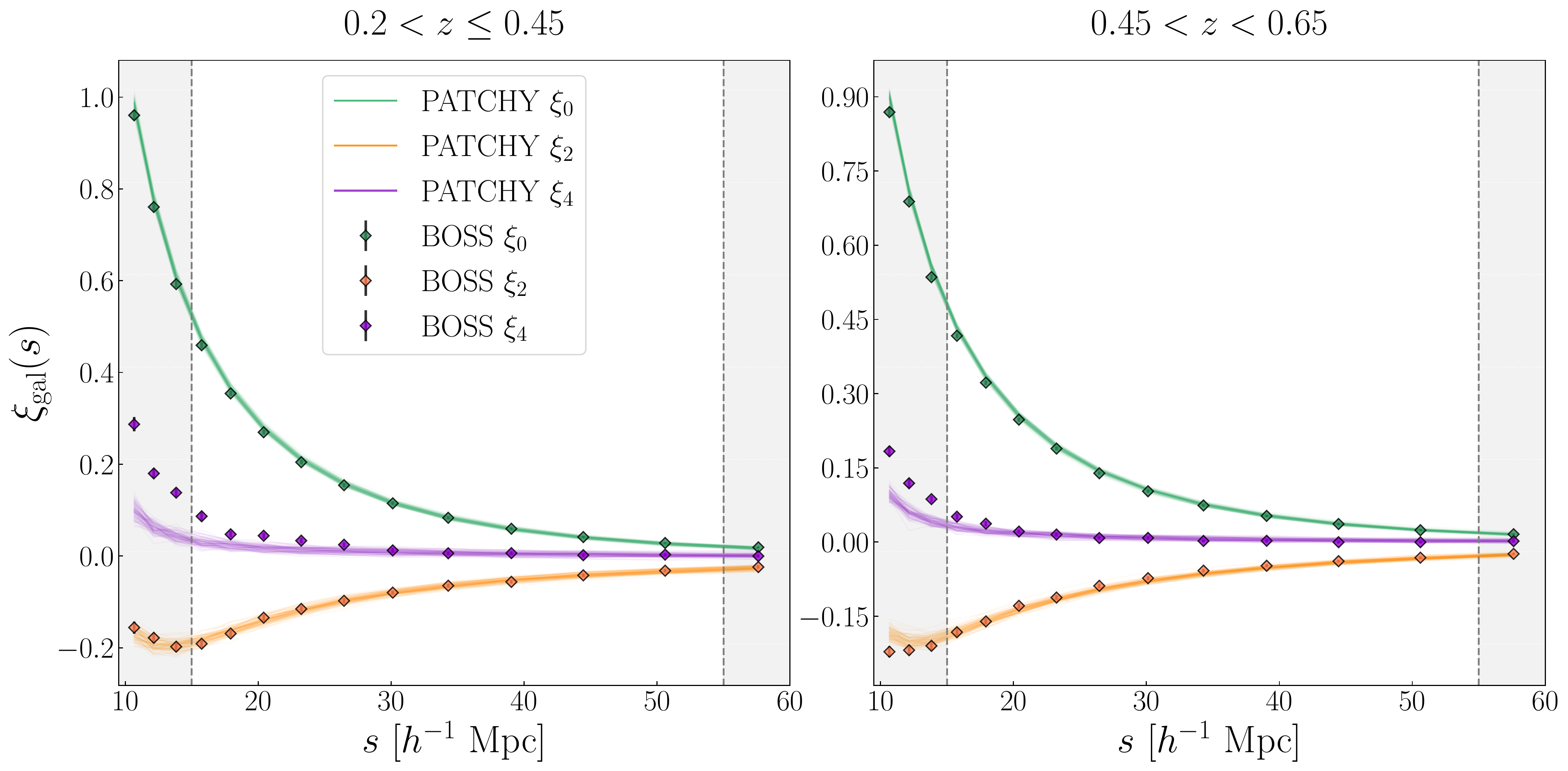}
\caption{MultiDark PATCHY and BOSS DR12 measurements of the redshift-space galaxy 2PCF multipoles, in the two redshift ranges considered in this analysis: $0.2 < z \leq 0.45$ (\textit{left}) and $0.45 < z < 0.65$ (\textit{right}). Solid lines show the data obtained from the $100$ PATCHY mock catalogs, while round markers indicate the measures from the BOSS data set. The correlation function monopole ($\xi_0$), quadrupole ($\xi_2$) and hexadecapole ($\xi_4$) are represented in green, orange and violet, respectively. The shaded regions outside the range $15 < s \,[h^{-1} \ \mathrm{Mpc}] < 55$ are those excluded from the statistical analysis.}\label{fig:multipoles}
\vspace{0.2cm}
\end{figure*}

\begin{figure*}[h]
\centering
\includegraphics[width=\textwidth]{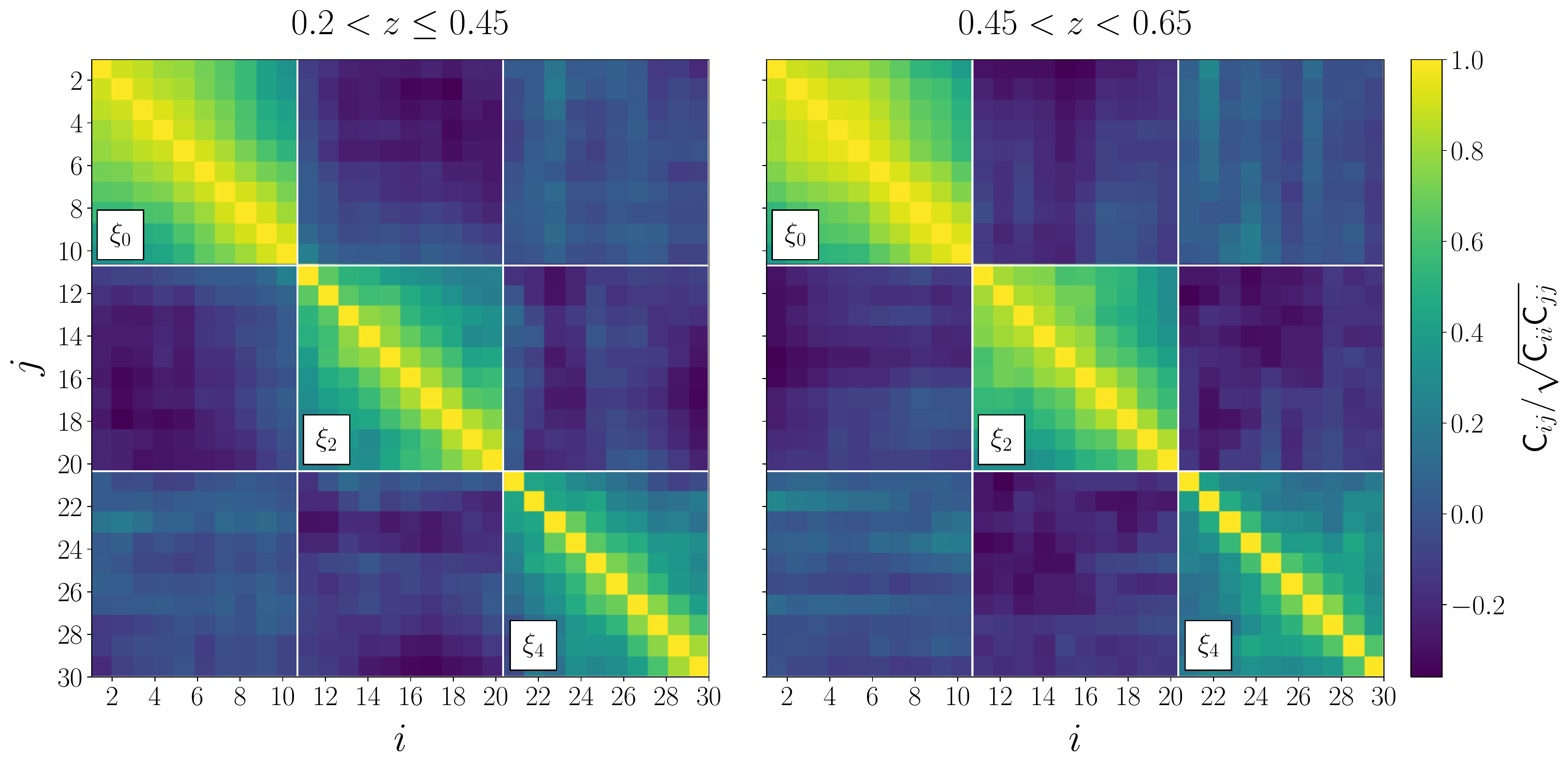}
\caption{Covariance matrix (normalized by its diagonal components) of the MultiDark PATCHY redshift-space galaxy 2PCF multipoles, for the two redshift selections considered in this analysis: $0.2 < z \leq 0.45$ (\textit{left}) and $0.45 < z < 0.65$ (\textit{right}). The galaxy separation bins $(i,j)$ are divided by vertical and horizontal white lines to separate the measures of the monopole ($\xi_0$), quadrupole ($\xi_2$) and hexadecapole ($\xi_4$) moments of the correlation function.}\label{fig:multipoles_cov}
\vspace{0.2cm}
\end{figure*}

We expand the 2D 2PCF into its multipole moments:
\begin{equation}
    \xi_\ell (s) \equiv \frac{2 \ell +1}{2} \int^{1}_{-1} \diff \mu \, \xi(s,\mu) L_\ell(\mu) \, ,
\end{equation}
where $\ell$ is the degree of the Legendre polynomials $L_\ell(\mu)$.
We consider the first three non-null multipole moments, i.e. the monopole, $\xi_0(s)$, the quadrupole $\xi_2(s)$, and the hexadecapole, $\xi_4(s)$:
\begin{equation}
 \xi(s,\mu) = \xi_0(s)L_0(\mu) + \xi_2(s)L_2(\mu) + \xi_4(s)L_4(\mu) \, ,
\end{equation}
where $L_0=1$, $L_2=(3\mu^2-1)/2$, $L_4=(35\mu^4-30\mu^2+3)/8$.

We report the results of these measures in \Cref{fig:multipoles}, for both the $100$ PATCHY mock realizations and BOSS catalogs. We notice the excellent agreement between the clustering properties of mocks and real data in each of the redshift ranges considered. Minor mismatches are found for the hexadecapole moment, although these deviations are significant only at small spatial scales, which are not considered in the following modeling of the analysis.

We use the $100$ measures of the two-point correlation multipoles performed on the PATCHY mocks to estimate the covariance matrix as:
\begin{equation}
    \hat{\textsf{C}}_{ij} = \frac{1}{N_s-1} \sum_{k=1}^{N_s} (y_i^k-\overline{y}_i)(y_j^k-\overline{y}_j) \, ,
\end{equation}
where $y=(\xi_0, \xi_2, \xi_4)$ is the combined vector of monopole, quadrupole and hexadecapole, $\overline{y}$ is the mean value computed by averaging the measures from the different realizations, $k$ runs over the different mock realizations, $N_s=100$, and $(i,j)$ define the galaxy separation bins.
We show the results for the two redshift sub-samples considered in \Cref{fig:multipoles_cov}. The correlation is found to be statistically relevant between the different separation bins, though almost negligible between the three 2PCF multipoles.

We adopt this covariance matrix in the Bayesian analysis to model the 2PCF multipoles in order to constrain the galaxy effective bias. We assume a standard Gaussian likelihood defined as follows:
\begin{equation} \label{eq:gaussian_lik}
    -2 \ln \mathcal{L} = \sum_{i,j=1}^{N} (y_i^{\rm d}-y_i^{\rm m}) \, \textsf{C}^{-1}_{ij} \, (y_j^{\rm d}-y_j^{\rm m}) \, ,
\end{equation}
where $N$ is the number of separation bins at which the multipole moments are computed, $\textsf{C}^{-1}$ the precision matrix, i.e. the inverse of the covariance matrix, and superscripts refer to data (${\rm d}$) and model ($\rm m$).

To model the multipoles of the redshift-space 2PCF we apply the approximation introduced by \cite{Taruya2010}, which has been extensively tested and validated in the literature \citep[see e.g.][]{delaTorre2012,Pezzotta2017,jorge2020}.
According to this model, the redshift-space dark matter power spectrum can be expressed as follows:
\begin{equation}\label{eq:TNS}
\begin{split}
    P^s(s,\mu) & = D(k, f, \mu, \sigma_{\rm v}) [ b^2 P_{\delta\delta}(k) \\
    & + 2 f b \mu^2 P_{\delta\theta}(k)+f^2\mu^4 P_{\theta\theta}(k) \\
    & + c_\mathrm{A}(k, \mu, f, b)+c_\mathrm{B}(k, \mu, f, b)] \, ,
\end{split}
\end{equation}
where $f$ is the linear growth rate, $b$ is the linear bias, $\sigma_{\rm v}$ is the pairwise velocity dispersion and $D$ is a damping factor that characterizes the random peculiar motions at small scales, assumed in this case to be Gaussian, i.e. $D(k, f, \mu, \sigma_{\rm v}) = \exp [-k^2f^2\mu^2\sigma_{\rm v}^2]$ \citep{davis1983,Fisher1994,zurek1994}. Then $P_{\delta\delta}(k)$ is the real-space matter power spectrum, $P_{\delta\theta}(k)$ and $P_{\theta\theta}(k)$ are the real space density-velocity divergence cross-spectrum and the real-space velocity divergence auto-spectrum, respectively. Lastly, in \Cref{eq:TNS}, $c_\mathrm{A}$ and $c_\mathrm{B}$ represent two additional terms to correct for systematic effects at small scales, which are computed according to \cite{Taruya2010} and \cite{delaTorre2012}, and can be expressed as a power series expansion of $b$, $f$ and $\mu$ \citep[see Appendix A of][for the details about the calculation of these correlation terms]{Taruya2010}.

We estimate the power spectra $P_{\delta\delta}(k)$, $P_{\delta\theta}(k)$ and $P_{\theta\theta}(k)$ in the standard perturbation theory, i.e. by expanding these statistics as sums of infinite terms, each one corresponding to a $n$-loop correction \citep[see e.g.][]{GilMarin2012}. Following \cite{jorge2020} and \cite{Marulli2021}, we consider in our analysis the corrections up to the $1$-loop order, computing the terms of the one-loop contribution with the \texttt{CPT Library}\footnote{\url{http://www2.yukawa.kyoto-u.ac.jp/~atsushi.taruya/cpt_pack.html}} \citep[see also][]{Bernardeau2002,GilMarin2012}.

The power spectrum multipoles are finally calculated from \Cref{eq:TNS} as:
\begin{equation}
    P_\ell(k) = \frac{2\ell +1}{2} \int^1_{-1} \diff \mu P^s (k,\mu)L_\ell(\mu) \, ,
\end{equation}
which is translated into configuration space by means of the following equation:
\begin{equation}
    \xi_\ell(s) = i^\ell \int^\infty_{-\infty} \frac{2k}{2\pi}k^2 P_\ell(k) j_\ell (k s) \, ,
\end{equation}
where $j_\ell$ are the spherical Bessel functions of order $\ell$. We model the clustering multipole moments at scales $15 < s \ [h^{-1} \ \mathrm{Mpc}] < 55$, estimating the linear bias of the galaxy samples considered in this work.

We assign wide positive priors to the free parameters of the \cite{Taruya2010} model, i.e. $b(z) \, \sigma_8(z)$, $f(z) \, \sigma_8(z)$ and $\sigma_\mathrm{v}$, maximizing their posterior distributions through a Bayesian analysis, on the $100$ PATCHY mock galaxy catalogs. We compute the average and the standard deviation of the best-fit values obtained with PATCHY mocks to estimate $b_\mathrm{eff} \, \sigma_8$ and its uncertainty, respectively, for both the selected redshift sub-samples.
We apply the same modeling procedure to the BOSS DR12 galaxies, performing in this case an MCMC analysis to sample the full posterior distribution of the free model parameters, marginalizing over $f(z) \, \sigma_8(z)$ and $\sigma_\mathrm{v}$, obtaining the values of $b_\mathrm{eff} \, \sigma_8$ reported in \Cref{sec:model_calibration}.

\section{Method validation} \label{appendix:PATCHY_constraints}

\begin{figure*}
\centering
\hspace{0.5cm}
\includegraphics[width=0.47\textwidth]{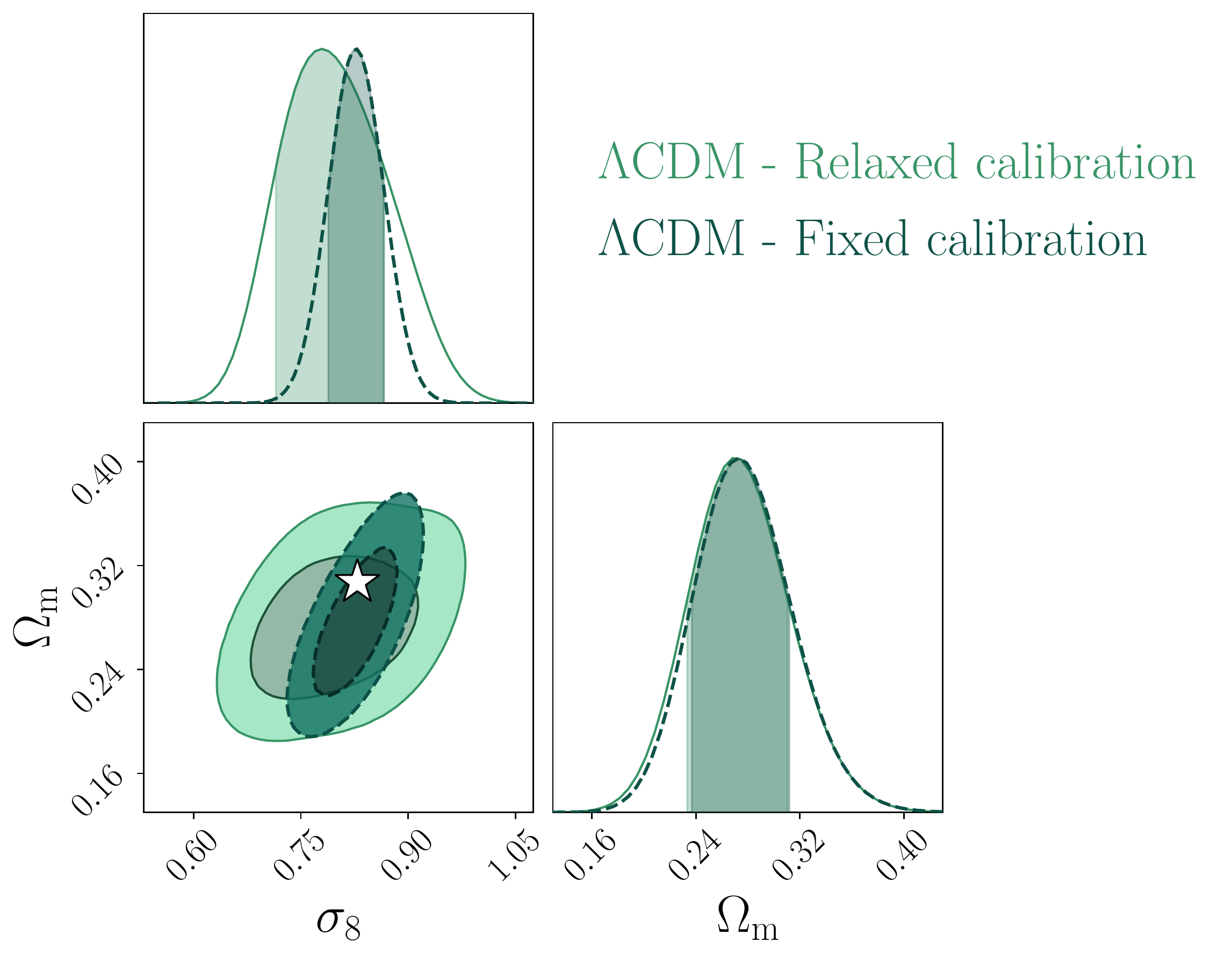}
\includegraphics[width=0.47\textwidth]{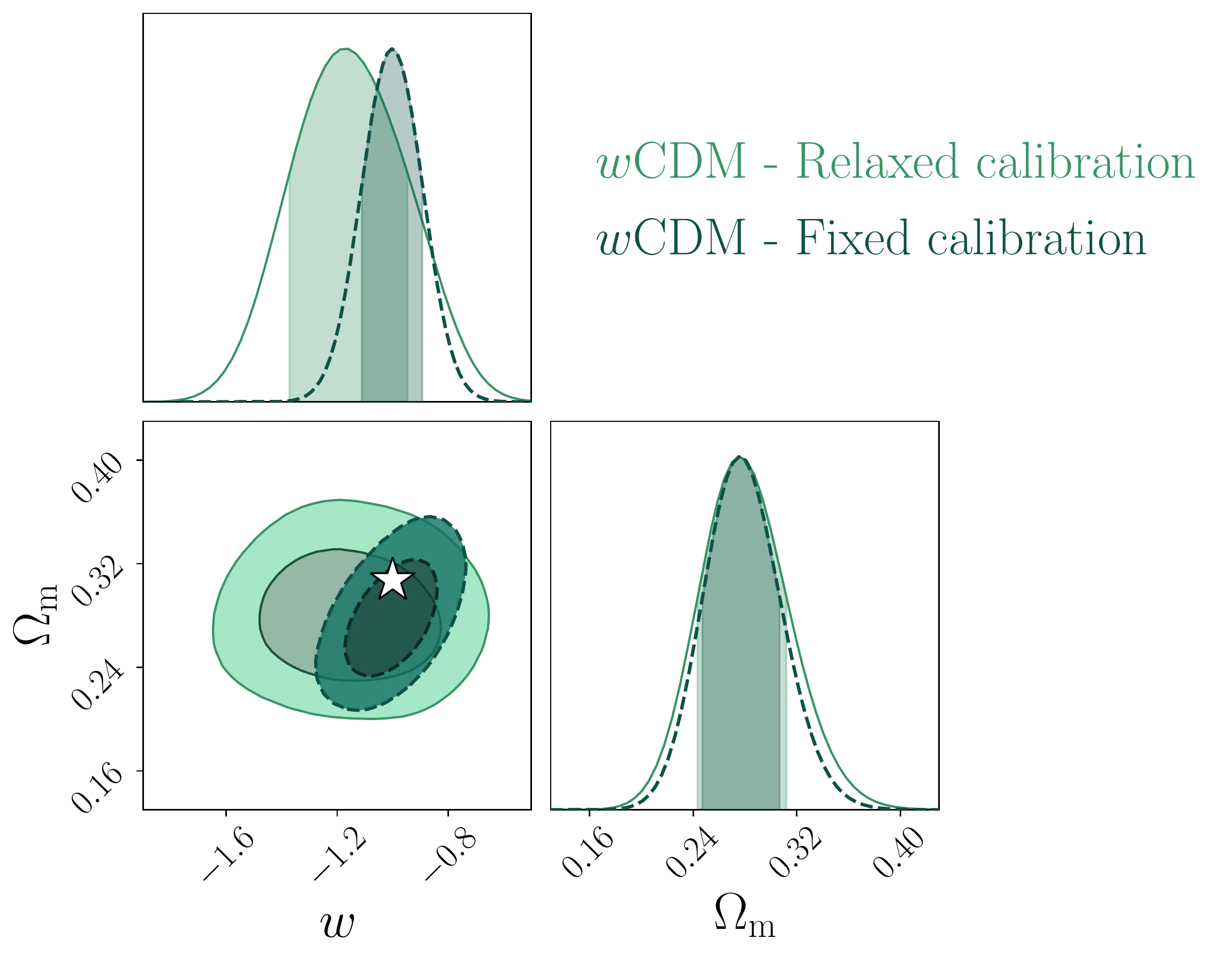}
\caption{As in \Cref{fig:BOSS_constraints}, but for the test constraints obtained from the size function of voids identified in the $100$ mock catalogs of the MultiDark PATCHY simulations. Dark and light green colors indicate the results obtained by fixing and relaxing the calibration of the parameters $C_\mathrm{slope}$ and $C_\mathrm{offset}$, respectively. With a white star we show the input cosmological parameters of the simulation. \label{fig:PATCHY_constraints}}
\end{figure*}

\begin{table*}
    \centering
    \caption{Maximum posterior values and $68\%$ uncertainties associated with the confidence contours reported in \Cref{fig:PATCHY_constraints}. The description is analogous to that in \Cref{tab:BOSS_results}, but for the test constraints obtained from the MultiDark PATCHY mocks.} \label{tab:PATCHY_results}
    \begin{tabular}{ccccccccccc}
        \toprule
		\multirow{2}{*}{\shortstack{ \\ \\ Model \\ \\ calibration}} & & \multicolumn{4}{c}{$\Lambda$CDM} & & \multicolumn{4}{c}{$w$CDM} \\
		\noalign{\vspace{0.1cm}}
		\cline{3-6}
		\cline{8-11}
		\noalign{\vspace{0.1cm}}
		& & $\Omega_\mathrm{m}$ & $\sigma_8$ & $C_\mathrm{slope}$ & $C_\mathrm{offset}$ & & $\Omega_\mathrm{m}$ & $w$ & $C_\mathrm{slope}$ & $C_\mathrm{offset}$\\ 
		\noalign{\vspace{0.1cm}}
		\hline
		\noalign{\vspace{0.1cm}}
		Relaxed & & $0.268^{+0.042}_{-0.035}$ & $0.779^{+0.087}_{-0.064}$ & $-1.92^{+0.55}_{-0.47}$ & $4.69^{+0.67}_{-0.76}$ & & $0.276^{+0.036}_{-0.032}$ & $-1.16^{+0.22}_{-0.21}$ & $-2.02^{+0.66}_{-0.61}$ & $4.74^{+0.87}_{-0.82}$ \\
		\noalign{\vspace{0.1cm}}
		Fixed & & $0.273^{+0.039}_{-0.036}$ & $0.828^{+0.038}_{-0.040}$ & $-1.62$ & $4.30$ & & $0.275^{+0.031}_{-0.028}$ & $-1.00\pm 0.11$ & $-1.62$ & $4.30$ \\ 
		\noalign{\vspace{0.1cm}}
		\hline
		\bottomrule
    \noalign{\vspace{0.3cm}}
    \end{tabular}
\end{table*}

Here we validate our data analysis method on the MultiDark PATCHY simulations using the void size function model calibrated in \Cref{sec:model_calibration}. This exercise is meant to check the validity of the methodology we introduced in this work and to quantify the impact of the survey characteristics on the resulting cosmological constraints from void counts.

We apply the Bayesian analysis described in \Cref{sec:cosmological_modeling} on the average void number counts reported in \Cref{fig:PATCHY_VSF}, considering both the \textit{fixed} and \textit{relaxed calibration} approaches. \Cref{fig:PATCHY_constraints} shows the results for the two cosmological models tested in this work, i.e. the flat $\Lambda$CDM and $w$CDM model. In both scenarios we recover the true cosmological parameters of the simulation within $68\%$ of uncertainties, confirming the reliability of the adopted pipeline. 

As expected, the orientation and shape of the confidence contours are analogous to those presented in \Cref{fig:BOSS_constraints} for the BOSS galaxies. 
The results we present in \Cref{tab:PATCHY_results} are compatible with those reported in \Cref{tab:BOSS_results}, although with smaller uncertainties. By comparing these results we notice that the main difference is in the precision on $\Omega_{\rm m}$, which improves by a factor $\sim 1.7$. 
However, this is due not only to the change of the volume coverage between BOSS and PATCHY galaxy catalogs. By comparing the square root of the effective volumes explored by \texttt{VIDE} (see \Cref{sec:data_prep}), we find that the two samples merely differ by a factor of $\sim 1.33$.
We also have to consider the higher galaxy number density and the more extended void radius range covered by the PATCHY samples with respect to the BOSS data set (see \Cref{fig:VIDE_VSF}). These specifics play a role in further improving the constraining power of the void size function, by increasing the statistical relevance of the void catalog and extending the spatial scales on which the Vdn model is sampled. This result highlights the importance of taking into account the impact of both the volume coverage and the tracer number density on void numbers for the design of future redshift surveys.

Despite the different characteristics of the PATCHY catalogs leading to an overall increase in the precision on the cosmological parameters with respect to the BOSS data, we note that part of the constraining power is absorbed by the Vdn model nuisance parameters $C_\mathrm{slope}$ and $C_\mathrm{offset}$, when not kept fixed to their best-fit values. This shows how a rigorous calibration of these parameters is fundamental to maximize the effectiveness of the void size function as cosmological probe \citep{Contarini2020,Verza2022}.

\section{Systematic errors} \label{appendix:systematics}
In this appendix, we evaluate the impact of the main systematic uncertainties possibly affecting our cosmological constraints by repeating the analysis on the PATCHY void counts considering different assumptions. We will focus on the \textit{relaxed calibration} approach only, as it provides the most reliable constraints.

\subsection{Calibration uncertainties}\label{sec:calib_sys}
We perform here the analysis on the PATCHY mocks by increasing the statistical errors of the void number counts. Specifically, during the calibration procedure, we associate to void counts Poissonian errors multiplied by a factor of $2$, $3$ and $4$. This approach is aimed at reproducing the effect of calibrating the void size function using mock catalogs with a volume $4$, $9$ and $16$ times smaller, respectively.

\begin{figure}
\centering
\hspace{-0.4cm}
\includegraphics[height=0.24\textwidth]{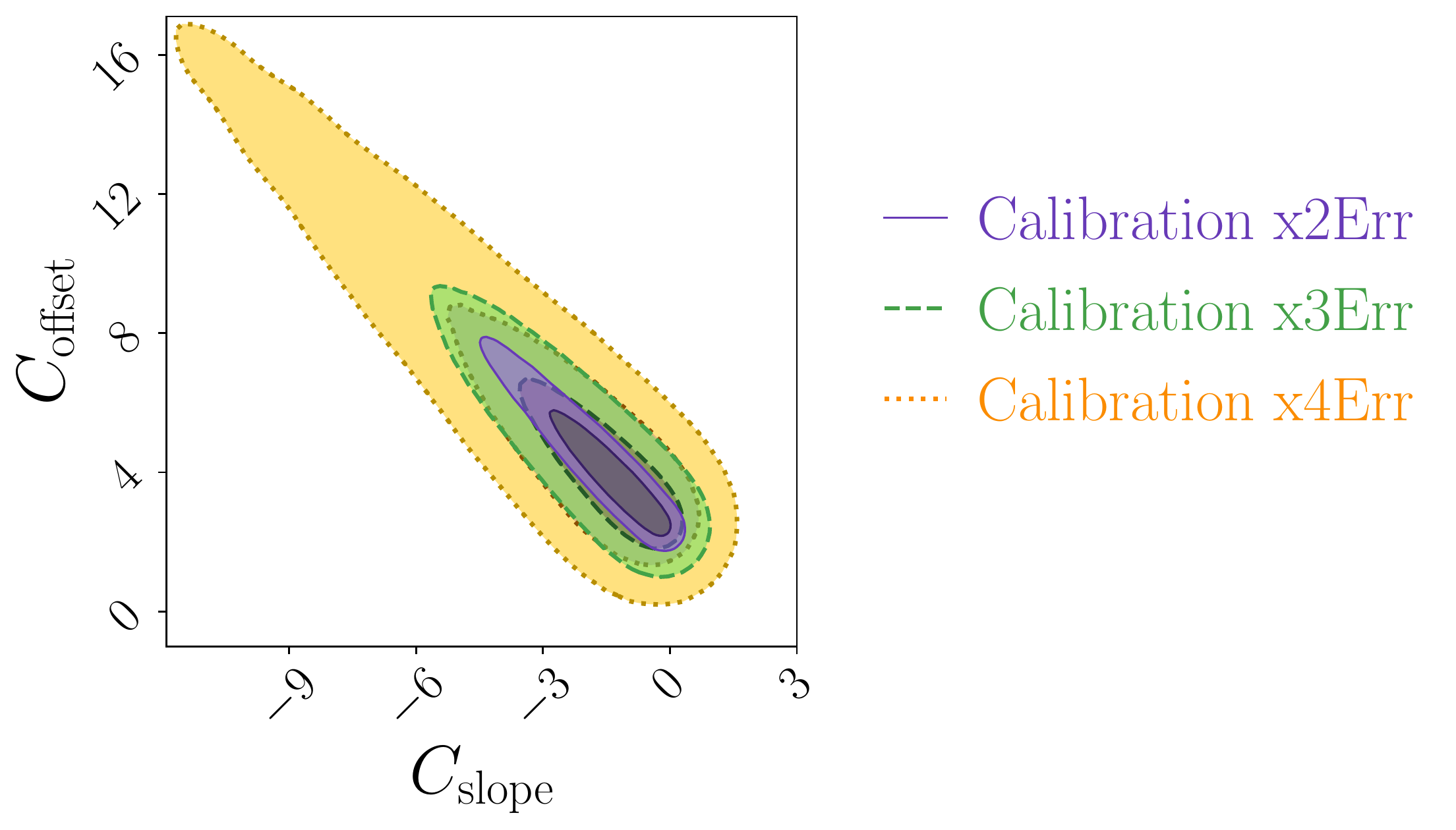}
\includegraphics[height=0.24\textwidth]{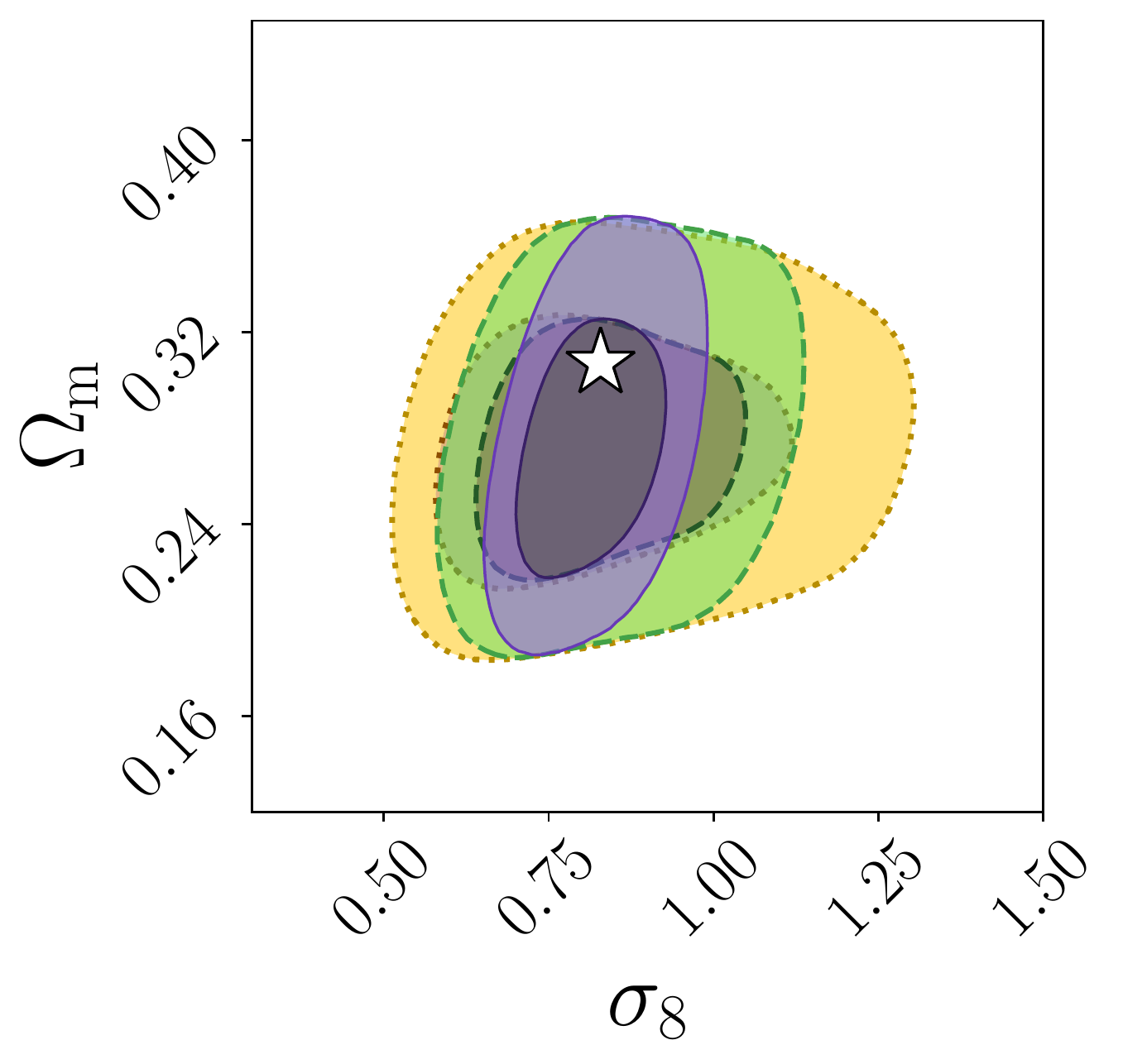}
\includegraphics[height=0.242\textwidth]{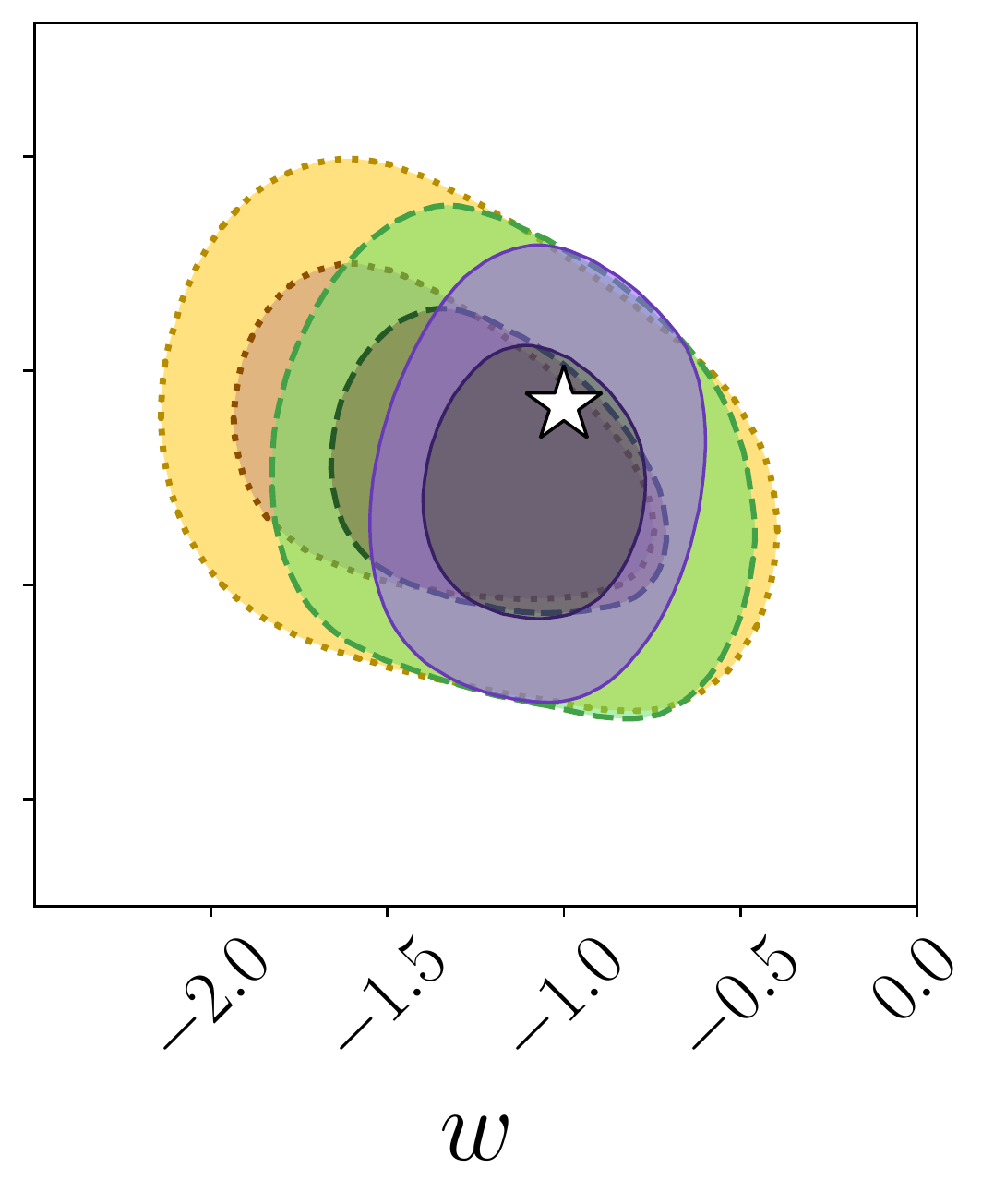}
\caption{$68\%$ and $95\%$ confidence contours obtained from the modeling of PATCHY void counts. \textit{Top}: confidence contours in the $C_{\rm slope}$--$C_{\rm offset}$ parameter plane, computed by increasing the void count uncertainty by a factor of $2$, $3$ and $4$. \textit{Bottom}: confidence contours in the parameter planes $\Omega_{\rm m}$--$\sigma_8$ (\textit{right}) and $\Omega_{\rm m}$--$w$ (\textit{left}). These contours are derived using the distributions shown in the top panel as 2D priors for the calibration parameters. The white star indicates the true cosmological values of the PATCHY simulations.} \label{fig:calibrationerr}
\vspace{0.1cm}
\end{figure}

In the upper panel of \Cref{fig:calibrationerr} we show the 2D posterior distribution of the parameters $C_\mathrm{slope}$ and $C_\mathrm{offset}$, for the three analyzed cases. We note that the enlargement of the confidence contours is mainly restricted to the degeneracy direction of the two calibration parameters.
In the lower panels of \Cref{fig:calibrationerr} we show instead the test constraints obtained for the parameter planes $\Omega_{\rm m}$--$\sigma_8$ ($\Lambda$CDM scenario) and $\Omega_{\rm m}$--$w$ ($w$CDM scenario).
In all the presented cases, the true values of the simulation cosmological parameters remain in the $68\%$ confidence region of the posterior distributions. 
As expected from the comparison between the \textit{relaxed calibration} and \textit{fixed calibration} methods (see \Cref{fig:PATCHY_constraints}), the impact of the nuisance parameter prior on $\Omega_\mathrm{m}$ is almost negligible, especially in the $\Lambda$CDM case.
In fact, the relative error on $\Omega_\mathrm{m}$ is stable around $14\%$. The effect on the constraints on $\sigma_8$ and $w$ is instead more important because of their degeneracy with the calibration parameters. The relative error on $\sigma_8$ increases from $9\%$ to $23\%$, when increasing the calibration errors from $2$ to $4$ times the Poissonian error. Analogously, the relative error on $w$ increases from $18\%$ to $25\%$.
However, we underline that the effects shown in this test derive from extreme calibration scenarios.

\subsection{Fiducial cosmology}\label{sec:fiducial_sys}

In this section we repeat the whole analysis, assuming a fiducial cosmology which is different from that of Planck2013. We run the void finder and the cleaning procedure on the $100$ PATCHY catalogs, considering a value of $\Omega_\mathrm{m}$ either decreased or increased by three times the standard deviation reported in \cite{Planck2018_results}.
Specifically, for this test we assume $\Omega_\mathrm{m}=0.279$ ($-3\sigma$) and $\Omega_\mathrm{m}=0.335$ ($+3\sigma$), modifying the value of $\Omega_\mathrm{de}$ to keep the Universe flatness. Moreover, we highlight that the accepted void radius range is recomputed during the analysis according to the value of ${\rm MGS}(z)$ in the considered cosmology.

\begin{figure}
\centering
\includegraphics[height=0.252\textwidth]{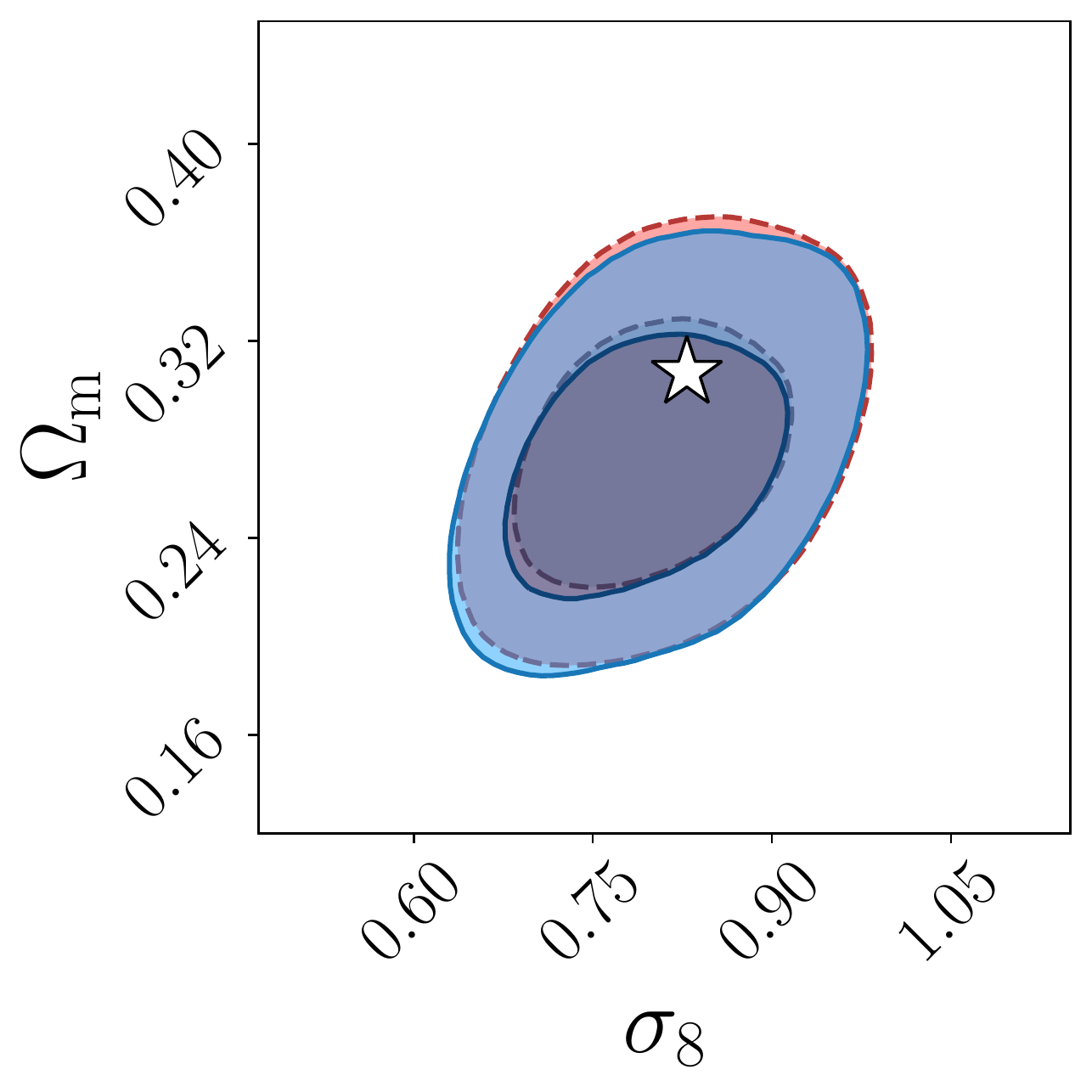}\hfill
\includegraphics[height=0.254\textwidth]{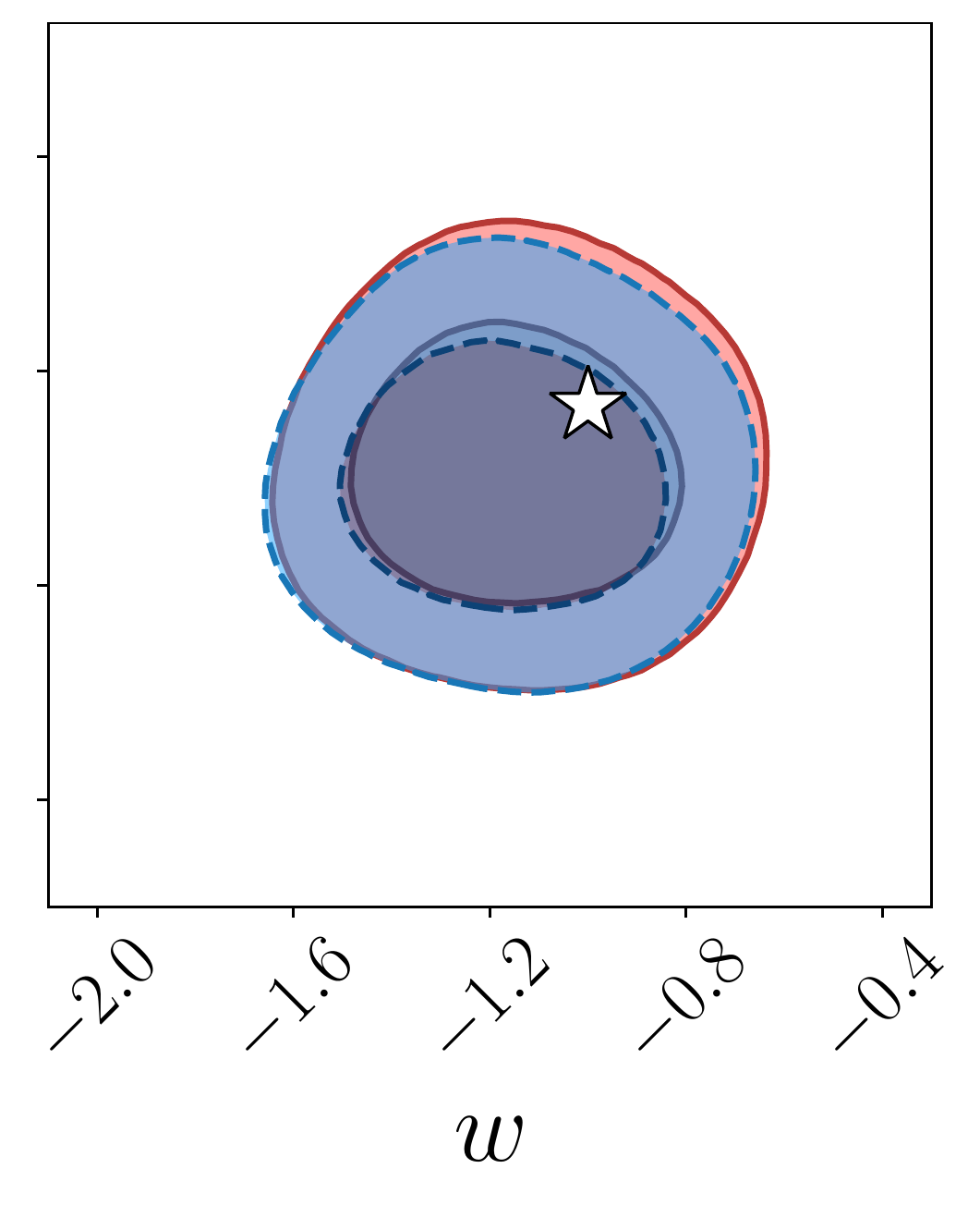}
\caption{$68\%$ and $95\%$ confidence contours computed by assuming different fiducial cosmologies in the modeling of PATCHY void counts, for the $\Lambda$CDM cosmology ($\Omega_{\rm m}$--$\sigma_8$ plane, \textit{right}) and the $w$CDM ($\Omega_{\rm m}$--$w$ plane, \textit{left}). 
In blue and red we report the results derived for a flat cosmological model with $\Omega_\mathrm{m}=0.279$ and $\Omega_\mathrm{m}=0.335$, respectively. We indicate with a white star the true cosmological parameters of the PATCHY simulations.} \label{fig:fiducialcosmology}
\vspace{0.1cm}
\end{figure}

We show in \Cref{fig:fiducialcosmology} the 2D posterior distributions obtained for the parameter spaces $\Omega_{\rm m}$--$\sigma_8$ and $\Omega_{\rm m}$--$w$. The two confidence contours are statistically compatible with each other, and the true cosmological values of the PATCHY simulations are recovered within the $68\%$ of uncertainty in both cases.

We report in \Cref{tab:fiducialCosm_results} the corresponding marginalized constraints computed, for the $\Lambda$CDM and $w$CDM scenarios. We note that, especially when exploring fiducial $\Omega_\mathrm{m}$ values lower than the true one (see the $\Omega_\mathrm{m}=0.279$ case), the underestimation of these cosmological parameters starts to become statistically significant. This issue will be analyzed in detail in future works.

\begin{table}
    \centering
    \caption{Maximum posterior values and $68\%$ uncertainties of the cosmological constraints computed assuming $\Omega_\mathrm{m}=0.279$ and $\Omega_\mathrm{m}=0.335$ as fiducial. We report the results for both the $\Lambda$CDM (\textit{upper} rows) and $w$CDM (\textit{lower} rows) scenarios. The associated confidence contours are shown in \Cref{fig:fiducialcosmology}.} \label{tab:fiducialCosm_results}
    \begin{tabular}{cccc}
        \toprule
        \noalign{\vspace{0.01cm}}
		$\Lambda$CDM & & $\Omega_\mathrm{m}$ & $\sigma_8$ \\
		\noalign{\vspace{0.1cm}}
		\hline
		\noalign{\vspace{0.1cm}}
		fiducial $\Omega_\mathrm{m}=0.279$ & & $0.265^{+0.041}_{-0.034}$ & $0.775^{+0.089}_{-0.064}$ \\
		\noalign{\vspace{0.1cm}}
		fiducial $\Omega_\mathrm{m}=0.335$ & & $0.270^{+0.042}_{-0.034}$ & $0.786^{+0.084}_{-0.067}$ \\ 
		\noalign{\vspace{0.1cm}}
		\hline
  		\noalign{\vspace{0.1cm}}
  		$w$CDM & & $\Omega_\mathrm{m}$ & $w$ \\
		\noalign{\vspace{0.1cm}}
		\hline
		\noalign{\vspace{0.1cm}}
		fiducial $\Omega_\mathrm{m}=0.279$ & & $0.278^{+0.035}_{-0.033}$ & $-1.18^{+0.22}_{-0.21}$ \\
		\noalign{\vspace{0.1cm}}
		fiducial $\Omega_\mathrm{m}=0.335$ & & $0.284^{+0.035}_{-0.036}$ & $-1.17^{+0.24}_{-0.20}$ \\ 
  		\noalign{\vspace{0.1cm}}
		\hline
		\bottomrule
    \end{tabular}
\end{table}

\subsection{Void radius selection}\label{sec:minR_sys}
As a final test on systematic errors, we focus here on the void radius selection. The choice regarding the minimum acceptable void radius holds significant importance as it directly impacts the strength of the resulting constraints: the greater the number of small voids included in the analysis, the more stringent the constraints derived from the void number counts. However, it is equally crucial to exclude the spatial scales that are influenced by void count incompleteness. In fact, the resolution of the tracer catalog affects the identification of small voids, which may be lost because of the Poisson noise.

In this work, we restricted our cleaned void sample to radii larger than $2.5$ times the MGS of the galaxy catalog (see \Cref{sec:data_prep,sec:model_calibration}). This leads to a minimum void radius of about $R_{\rm min} = 36.37$ and $35.13 \ h^{-1} \ \mathrm{Mpc}$, for the redshift bins $0.2 < z \leq 0.45$ and $0.45 < z < 0.65$, respectively. For simplicity, we apply the same radius cut to both redshift bins, without relying on the MGS value. Here we consider three different radii selections: one slightly lower than the one used in the main analysis of this work, i.e. $R_{\rm min} = 35 \ h^{-1} \ \mathrm{Mpc}$, and two higher, i.e. $R_{\rm min} = 37.5$ and $40 \ h^{-1} \ \mathrm{Mpc}$.

\begin{figure}
\centering
\includegraphics[height=0.255\textwidth]{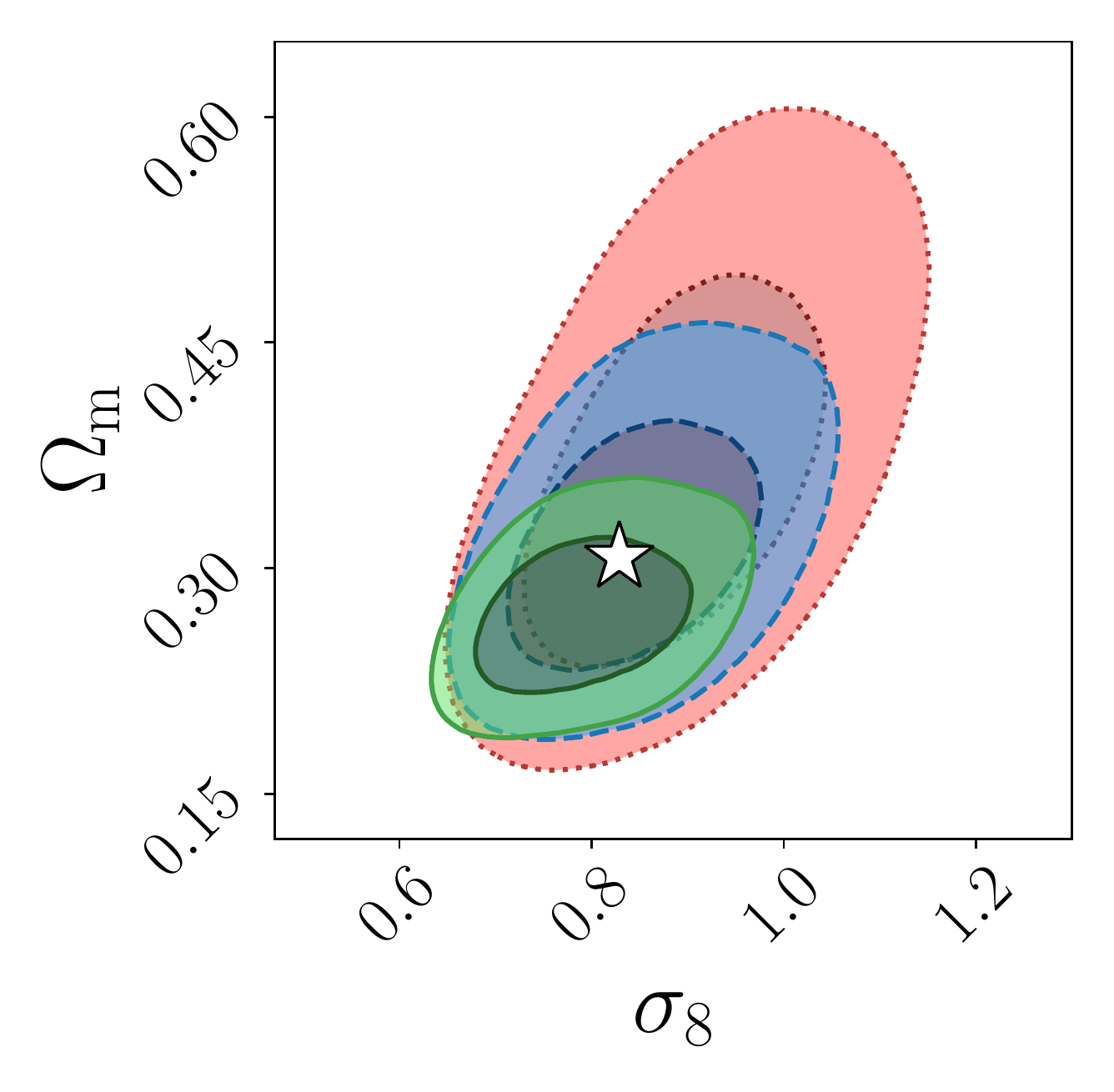}
\includegraphics[height=0.253\textwidth]{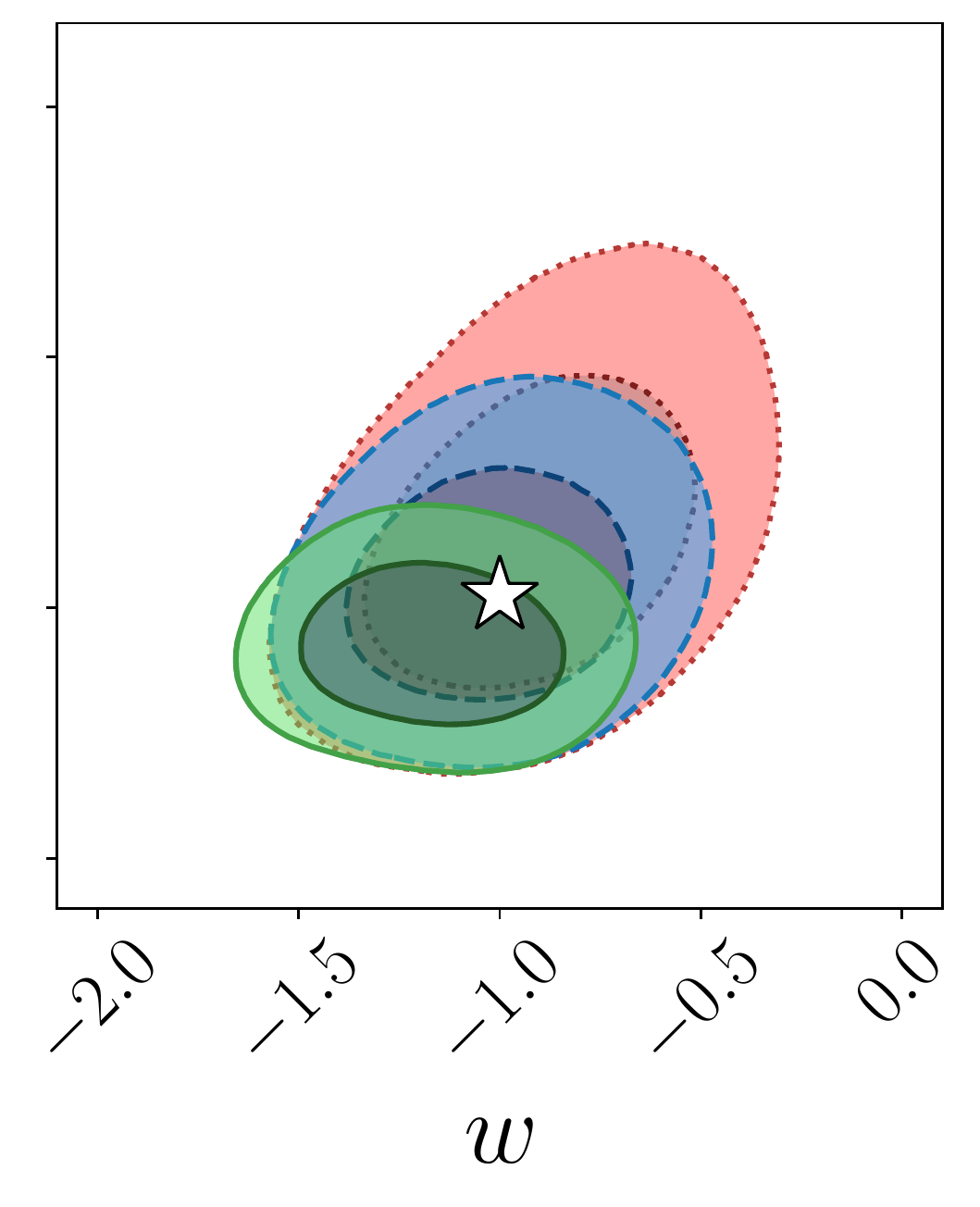}
\caption{As in \Cref{fig:fiducialcosmology}, but for different void radius selections. We represent in green, blue and red the test constraints derived by modeling the PATCHY void counts with $R_\mathrm{min}=35, \, 37.5, \, 40 \, \ h^{-1} \ \mathrm{Mpc}$, respectively. The first radius selection is slightly lower than the one used in the analysis presented in \ref{appendix:PATCHY_constraints}.}\label{fig:Rmin}
\vspace{0.1cm}
\end{figure}

In \Cref{fig:Rmin} we show the 2D posterior distributions computed for the $\Lambda$CDM and the $w$CDM scenarios by applying the presented radius selections. As expected, the cosmological constraints become weaker as the void minimum radius becomes larger. In the $\Lambda$CDM scenario, moving from the lowest to the highest radius cut, the relative error increases by a factor $1.9$ and $1.2$, for $\Omega_{\rm m}$ and $\sigma_8$, respectively. When considering the $w$CDM scenario, the increase is instead by a factor $1.6$, for both the parameters $\Omega_{\rm m}$ and $w$. However, it is important to emphasize that, in all three cases, the true cosmological parameters of the PATCHY simulations are within the $68\%$ confidence region.

\section{Parameter orthogonality} \label{appendix:orthogonality}

\begin{figure*}
\centering
\includegraphics[width=0.32\textwidth]{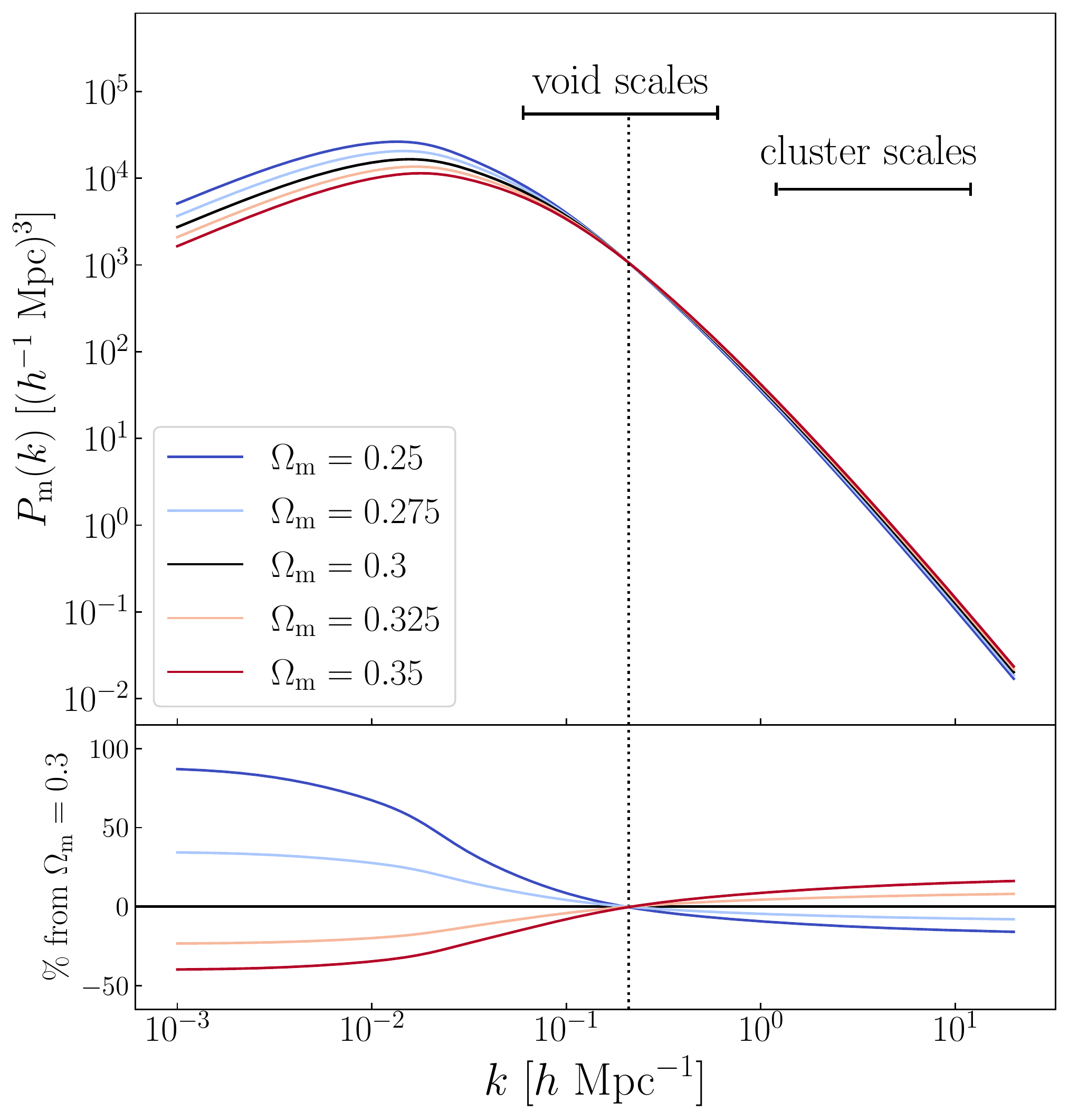}
\includegraphics[width=0.325\textwidth]{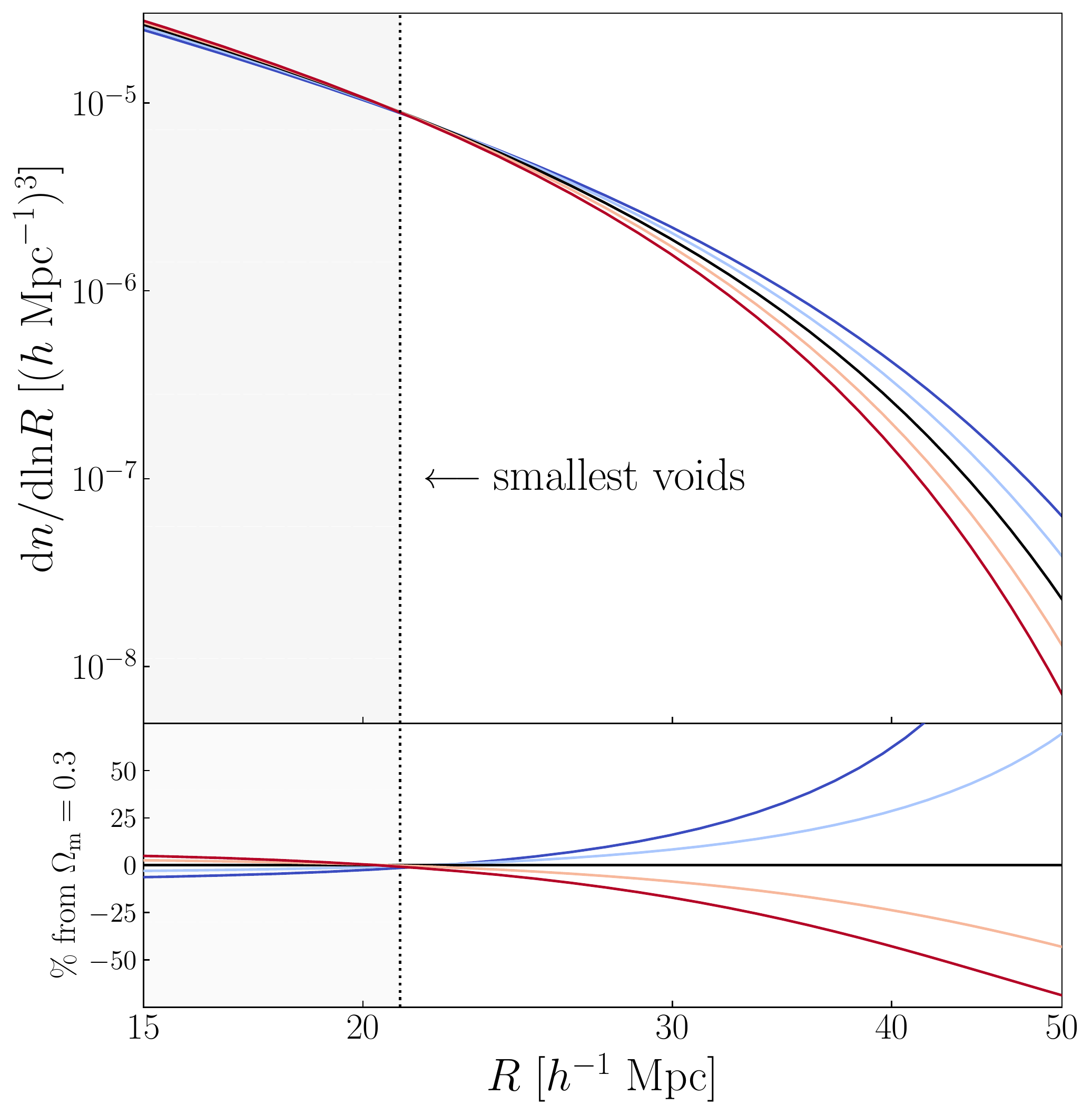}
\includegraphics[width=0.325\textwidth]{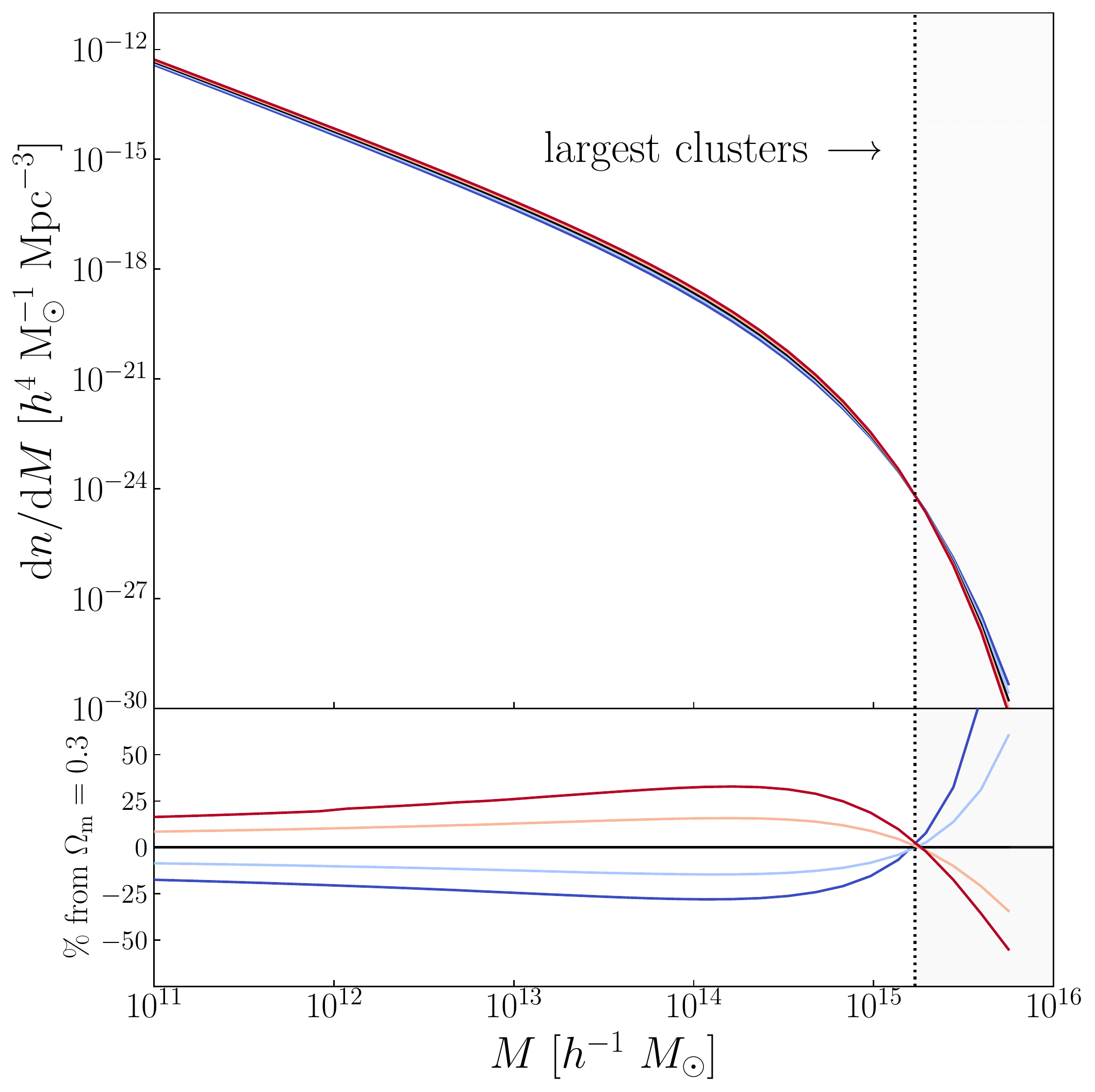}
\caption{Theoretical linear matter power spectrum (\textit{left}), void size function (\textit{center}) and halo mass function (\textit{right}) at $z=0.5$, computed with Planck2013 parameters \citep{Planck2013}, and normalized to have $\sigma_8=0.83$. We represent in different colors the results of varying $\Omega_{\rm m}$ and highlight with a dotted vertical line the pivot points that separate opposite trends. The bottom subplots show the residuals from the fiducial model with $\Omega_{\rm m}=0.3$. \label{fig:Pk}}
\vspace{0.2cm}
\end{figure*}

The orthogonality on the $\Omega_{\rm m}$--$\sigma_8$ parameter plane between the void size function and the standard probes can be understood based on the following observations. Let us first consider the effects of these cosmological parameters on the linear matter power spectrum. Its variance determines both the number density of voids and galaxy clusters, or similarly dark matter halos (see \Cref{sec:theory}).

Increasing the value of $\sigma_8$ is equivalent to shifting the normalization of the power spectrum at $z=0$ towards higher values, favoring the growth of matter perturbations. This enhances the formation of both galaxy clusters and cosmic voids. On the other hand, suppose we fix the matter power spectrum normalization $\sigma_8$, and increase the matter density of the universe, $\Omega_{\rm m}$. As shown in the left panel of \Cref{fig:Pk}, widely extended objects such as voids ($10 \lesssim r \ [h^{-1} \ \mathrm{Mpc}] \lesssim 100$) correspond to cosmic scales characterized by a negative power spectrum response to an increase in $\Omega_{\rm m}$. On the contrary, smaller cosmological objects, such as galaxy clusters ($0.5 \lesssim r \ [h^{-1} \ \mathrm{Mpc}] \lesssim 5$) are characterized by a positive response. The spatial scale that separates the two matter power spectrum trends in this plot corresponds to $k = 0.4 \ h \ \mathrm{Mpc}^{-1}$, but depends on multiple factors, e.g. the redshift and the cosmological model. Moreover, during their nonlinear evolution, clusters decrease and voids increase their comoving sizes by up to a factor of a few. However, this is not enough to revoke the difference in their Lagrangian scales when selected in Eulerian space.

Intuitively, for a given value of $\sigma_8$, so when the clustering strength is fixed, the formation of voids is disfavored with increasing $\Omega_{\rm m}$. This is because it makes large voids less underdense, while the collapse and accretion rate of overdensities is enhanced by the presence of more matter in the universe.
The counteractive behavior of these objects determines the complementarity of the constraints derived from the cluster mass function and the void size function \citep[see also][]{Pelliciari}. 

We substantiate this intuition in the center and right panels of \Cref{fig:Pk}, where we illustrate the effects of variations in $\Omega_{\rm m}$ on the void size function \citep{jennings2013} and the halo mass function \citep{PressSchechter}. It is evident how the trends of the two statistics are opposite in the typical radius and mass ranges: higher values of $\Omega_{\rm m}$ cause a decrease in the number of cosmic voids and an increase in the number of massive halos. Nevertheless, this behavior is different for very small voids and the most massive halos. In this example, the pivot points correspond to $R = 21 \ h^{-1} \ \mathrm{Mpc}$ for voids and $M = 1.7 \times 10^{15} \ h^{-1} \ M_\odot$ for halos. We note, however, that very small voids ($R \simeq 10 \ h^{-1} \ \mathrm{Mpc}$) are commonly not identified in wide-field surveys, because of their limited spatial resolution given by the tracer density. On the other hand, very massive clusters ($M \simeq 10^{15} \ h^{-1} \ M_\odot$) are statistically very rare and constitute a minor contribution to the mass function constraints. 

We finally underline that the reported statistics are just meant to portray a simplified view on the subject, and do not aim at reproducing any realistic distribution of cosmological objects. Indeed, a number of observational effects\footnote{For instance, the mass tracer bias and the AP correction affect the void size function, and the inclusion of the baryon component inside dark matter halos affects the halo mass function.} modify the theoretical predictions of these models, leading to a shift of the pivot scales.

\pagebreak

\bibliography{bibliography}{}
\bibliographystyle{aasjournal}

\end{document}